# Transfer of Freestanding Fluoropolymer Films for Advanced Semiconductor Devices

Mohammad Monish[1,*], Koki Hino[1,2], Yosuke Sasama[3], Masato Urakami[4], Takehiro Ota[4], Kenji Sakamoto[5]
Kenichiro Takakura[4], Yamaguchi Takahide[1,2,*]

[1]Research Center for Materials Nanoarchitectonics, National Institute for Materials Science, Tsukuba, Ibaraki 305-0044, Japan
[2]University of Tsukuba, Tsukuba, Ibaraki 305-8571, Japan
[3]International Center for Young Scientists, National Institute for Materials Science, Tsukuba, Ibaraki 305-0044, Japan
[4]National Institute of Technology, Kumamoto College, Koshi, Kumamoto 861-1102, Japan
[5]Research Center for Macromolecules and Biomaterials, National Institute for Materials Science, Tsukuba, Ibaraki 305-0047, Japan

[*]**Email:** MOHAMMAD.Monish@nims.go.jp; YAMAGUCHI.Takahide@nims.go.jp

## Abstract

High-quality dielectric films are essential for fabricating advanced electronic devices, but their direct deposition often degrades the films and their underlying interfaces, which compromises device performance, especially on sensitive or low-adhesion surfaces. To overcome these limitations, film transfer methods enable the integration of high-quality dielectric films onto such surfaces without damaging the underlying interfaces. However, existing transfer methods have predominantly focused on high-dielectric-constant (high-$\kappa$) materials, leaving a critical gap for transferable, high-quality low-$\kappa$ alternatives, which are required for enabling low-power and high-speed electronics. Herein, we address this need by demonstrating a method to integrate freestanding low-$\kappa$ fluoropolymer dielectric films with smooth surface morphology onto diverse substrates, including low-adhesion surfaces like hydrogen-terminated diamond. The transferred films revealed high breakdown fields of $8.0 \pm 1.2$ MV cm$^{-1}$, with leakage current density remaining typically below $10^{-7}$ A cm$^{-2}$ before the breakdown. The incorporation of these fluoropolymer films as gate dielectrics in $p$-channel hydrogen-terminated diamond field-effect transistors resulted in transfer and output characteristics with negligible hysteresis, high channel mobility ($\approx$400 cm$^2$ V$^{-1}$ s$^{-1}$) and a low interface trap density ($\leq 3 \times 10^{11}$ cm$^{-2}$ eV$^{-1}$). These findings highlight the versatility of the transfer method and position freestanding fluoropolymers as a promising platform for forming high-quality dielectric/semiconductor interfaces for advanced electronics.

## 1. Introduction

The continuing progress in unlocking new possibilities for high-performance, energy-efficient, and multifunctional semiconductor devices is paving the way for breakthroughs in power electronics, communication technologies, and wearable electronics.[1-3] These advancements have attracted great interest in the development of next-generation transistors and hybrid electronic systems[4,5] through the integration of high-quality gate dielectrics, which enable reduced leakage current, enhanced carrier modulation, and improved device stability. The dielectric films are routinely grown by atomic layer deposition (ALD), chemical vapor deposition (CVD), pulsed laser deposition (PLD) and sputtering; however, these techniques often involve high temperatures or energetic species, which can significantly damage chemically sensitive surfaces and degrade interfacial quality.[6-8] Furthermore, widely used dielectric films, such as $SiO_2$, $Al_2O_3$, and $HfO_2$, have demonstrated incompatibility with two-dimensional (2D) materials or surfaces lacking dangling bonds, resulting in poor adhesion and degraded dielectric film quality.[8-10] These challenges have spurred significant research efforts aimed at improving dielectric film deposition approaches[11-13] and developing alternative strategies to enable non-destructive dielectric integration. Among these emerging approaches, the transfer or lamination of pre-formed dielectric films has recently gained considerable attention because it avoids the limitations of direct deposition.[14-20] Although these methods have demonstrated improvements in device performance, they have predominantly focused on high-$\kappa$ materials. This, however, has left a critical gap for transferable, high-quality low-$\kappa$ alternatives, which are crucial for minimizing power consumption and parasitic capacitance in high-frequency applications.

CYTOP, an amorphous fluoropolymer, has potential to fill this gap, as it exhibits outstanding dielectric properties, including low-$\kappa$ of 2.0 - 2.1, high electrical resistivity, and a large breakdown strength.[21,22] In addition, CYTOP exhibits chemical inertness, extremely low water absorptivity (<0.01%), and can form highly uniform, large-area films via solution processing, enabling scalable and reliable device integration. Its high optical transparency (>95%) and low refractive index (1.34)[21] further broaden its applicability to optoelectronic and photonic devices. Owing to these properties, CYTOP has been employed in a wide range of applications, including organic field-effect transistors (OFETs), bioelectronics, anti-reflective coatings and photonic encapsulation.[23-28] Although the inherently low surface energy and chemical inertness of CYTOP film surface posed challenges in standard lithographic processing, recent advances in surface activation methods, resist formulation and adhesion-promoting treatments have enabled the successful implementation of high-resolution photolithography on CYTOP films.[29,30] Despite the excellent properties of CYTOP and the recent advancements, most studies reported to date have focused on spin-coated CYTOP films, while the properties and potential applications of pre-formed freestanding CYTOP dielectric films remain largely unexplored.

In this work, we demonstrate the transfer-integration of freestanding CYTOP films into semiconductor devices, without degrading the target surface, and thereby achieves high quality interfaces. The lamination of freestanding CYTOP films was achieved onto a wide range of substrates such as silicon, sapphire, diamond and CYTOP itself, enabling dielectric integration across diverse device architectures. The metal-insulator-metal (MIM) structures fabricated on sapphire substrate utilizing the transferred freestanding CYTOP dielectric films revealed the breakdown fields of $8.0 \pm 1.2$ MV $cm^{-1}$, while the leakage current densities remain substantially low ($<10^{-7}$ A $cm^{-2}$) before the breakdown. CYTOP film was also transferred to hydrogen-terminated (H-terminated) diamond, a dangling-bond free surface where weak adhesion characteristics and high sensitivity to the deposition environment often result in poor film quality and surface degradation. The *p*-channel H-terminated diamond FETs incorporating the transferred CYTOP gate dielectric were fabricated, exhibiting negligible hysteresis in output and transfer characteristics, along with a substantially low interface trap density and high hole mobility. These findings demonstrate the excellent properties of CYTOP and validate the feasibility of the present transfer method, thus establishing a versatile platform for dielectric integration and broadening the scope of CYTOP applications in advanced electronic and optoelectronic technologies.

## 2. Transfer and Characterization of Freestanding CYTOP Films

### 2.1. Fabrication of Freestanding CYTOP Films

The fabrication and transfer process for freestanding CYTOP fluoropolymer films is schematically illustrated in Figure 1 and the details are provided in the *Experimental Section*. Polyacrylic acid (PAA), a hydrophilic and water-soluble polymer,[31] was used as a sacrificial layer to enable the subsequent lift-off or exfoliation of the CYTOP film. The PAA layer was first spin-coated onto silicon (Si) substrates and then baked. Subsequently, type-S CYTOP[21] was spin-coated onto the PAA/Si and baked, forming the final stack structure illustrated in Figure 1(a). The baking process is provided in *Experimental Section*. The photograph of prepared CYTOP/PAA/Si sample is shown in Figure S1(a) of Supporting Information. To facilitate the exfoliation and the transfer of freestanding CYTOP film without any distortion, a high-temperature resistant Kapton tape was fixed onto CYTOP/PAA/Si sample with a pre-cut central aperture to define a localized area of interest for clean CYTOP film transfer, as shown in Figure 1(b) (see also Figure S1(b) for corresponding photograph). The Kapton provides mechanical support and minimizing wrinkle formation, while simultaneously enabling a feasible transfer of CYTOP films without contaminating the localized area of interest. After this step, the assembled stack (Kapton/CYTOP/PAA/Si) was floated on a deionized (DI) water bath, which was kept on a hot plate with temperature of 50 °C to accelerate the dissolution of the underlying PAA layer, as shown in Figure S1(c). Although PAA is water-soluble at room temperature, elevated temperature ensures faster and more consistent water penetration to dissolve PAA. During the dissolution of PAA, only the lower surface of CYTOP film was exposed to water, while the upper surface remained pristine thereby preserving the surface integrity and minimizing contamination on the intended transfer surface. After the complete dissolution of the PAA layer, the underlying Si substrate was detached and the Kapton-tape-supported CYTOP film remained floating on the water surface, as illustrated in Figure 1(c). To prevent wrinkle formation, enable accurate film alignment and improve handling during lamination, the tape-supported CYTOP film was mounted onto a copper frame with a circular aperture near one of the edges of the frame, as illustrated in Figure 1(d). The aperture in copper frame allows alignment of freestanding CYTOP film during the lamination under the optical microscope, as shown in Figures S1(d)-S1(f). Finally, the assembly was baked at 100 °C for 5 min to remove residual moisture and stored in a vacuum-sealed container to prevent contamination from airborne particulates.

### 2.2. Transfer of Freestanding CYTOP Films

The freestanding CYTOP films prepared as discussed above were transferred onto various substrates, including Si, CYTOP laminated Si, sapphire, and H-terminated diamond. Figure 1(e) depicts the lamination step of the copper-frame-supported CYTOP film onto the target substrate, while Figure 1(f) presents schematic of the final sample. The transfer was carried out by using a movable stage, which provided x-y-z control of the copper frame holding the freestanding CYTOP film under an optical microscope, enabling accurate positioning and controlled contact with the target substrate. The target substrate was mounted on an adjacent sample stage equipped with a vacuum chuck to securely hold the substrate, which also featured a heater to bake the substrate when necessary to remove surface adsorbates during CYTOP film transfer. The corresponding photograph of sample stage and movable stage holding the copper frame is shown in Figure 2(a). The copper frame is fixed in a way that the pristine CYTOP surface faces downward, allowing direct contact with the target substrate. Prior to CYTOP lamination, the sample stage temperature was set to 110 °C. The aperture in the copper frame, which is larger than the size of substrate, was aligned above the substrate using fine x-y adjustments (Figure 2(a)) under the optical microscope. After alignment, the copper frame was gradually lowered to bring the CYTOP film into contact with the target substrate surface. The CYTOP films were found to adhere immediately to the substrate; therefore, the film and substrate were aligned parallel before contact to prevent misalignment or wrinkles. The freestanding CYTOP films were transferred to various substrates, including H-terminated diamond surface, which typically resists the deposition of metals and insulators. Figure 2(b) shows the photograph of CYTOP film laminated on H-terminated diamond substrate in ambient air, which was captured after slightly retracting the copper frame (or moved in upward direction) in the z-direction after lamination, while the diamond substrate remained fixed on the stage. It is seen that the CYTOP film is bent around the edge of the diamond substrate, demonstrating substantial bonding between CYTOP and the diamond surface. After turning off the vacuum and releasing the diamond substrate from the sample stage, the z-axis of the copper frame holding stage was raised further. The diamond substrate remained attached to the CYTOP film and was lifted along with it, as shown in Figure 2(c). The optical microscope images of CYTOP films laminated on various substrates are shown in Figures 2(d)-2(g). Minor wrinkles observed in the transferred CYTOP films were effectively removed by post-annealing at 130 °C (not shown), likely due to the relaxation of residual mechanical stress generated during the transfer process. These observations suggest that CYTOP can be used as a transferable film that adheres to various substrates.

### 2.3. Surface Morphology

The atomic force microscopy (AFM) was performed on CYTOP films to study the surface morphology. The surface morphology plays a crucial role in determining the effectiveness of dielectric films in electronic devices, particularly by ensuring uniform film coverage and minimizing local electric field enhancements that could lead to early breakdown.[32,33] Figure 2(h) displays the AFM image of spin coated CYTOP film on PAA/Si over a scan area of 20 µm × 20 µm, which reveals substantially smooth and featureless surface morphology. The AFM measurements were performed at multiple locations on four-different samples prepared under similar conditions, yielding a root-mean-square (rms) roughness of 0.45 ± 0.03 nm. The height histogram in Figure 2(i) corresponding the AFM measurements of four different samples confirms negligible variation, indicating high uniformity and consistent roughness values. In order to compare, the AFM measurements were also performed on Si substrate, PAA coated on Si and CYTOP film coated on Si without PAA, which are shown in Figure S2 of Supporting Information. The surface morphology and roughness of CYTOP films were also found to be nearly the same when directly spin coated on Si without PAA, as shown in Figure S2, indicating that the films morphology does not change due to underlying layer. The high uniformity and low-roughness of CYTOP films make it suitable for the transfer method.

### 2.4. Electric Breakdown Strength

The electric breakdown strength of freestanding CYTOP films was evaluated by fabricating MIM structure. Initially, bottom electrodes were patterned on cleaned sapphire substrates using maskless laser lithography followed by the deposition of Ti/Pt (5 nm/5 nm) electrodes by e-beam evaporation. After bottom electrodes deposition, the freestanding CYTOP film was laminated in ambient air following the transfer method discussed above. The film thickness was controlled in the range of 25 to 221 nm by systematically adjusting both the CYTOP solution concentration and the spin-coating rotation speed. Finally, after the lamination, top electrodes (Al/Au: 50 nm/100 nm) were deposited through a metal shadow mask using thermal evaporator. Figure 3(a) and the inset of Figure 3(b) show the schematic and optical microscope image of MIM structure, respectively. Figure 3(b) presents the current-voltage (*I-V*) characteristics of MIM devices, with the negative voltage applied to the top electrode and current measured from the bottom electrode. The *I-V* measurements show that the devices consistently exhibited low leakage current densities, typically remaining below $10^{-7}$ A cm$^{-2}$, and in all cases not exceeding the $10^{-6}$ A cm$^{-2}$ range prior to dielectric breakdown. The breakdown electric field was measured to be in the range of 5.9 – 10.4 MV cm$^{-1}$. The substantially high breakdown fields obtained across the devices confirm the excellent insulating characteristics of laminated CYTOP films. Moreover, these breakdown field values are comparable to those of directly spin-coated thick CYTOP films,[22] which demonstrates that our transfer approach preserves the insulating property of CYTOP. To further evaluate the dielectric properties of the transferred films, capacitance-voltage (*C-V*) measurements were performed on diamond FETs employing transferred CYTOP as the gate dielectric. The extracted relative dielectric constant was 2.0, which is consistent with previously reported values for CYTOP,[21] and will be discussed in detail in the next section. Figure 3(c) shows the breakdown fields at a leakage current density of $10^{-5}$ A cm$^{-2}$ plotted against the relative dielectric constant, which includes the average value from all the devices measured in this work (excluding the MIM device with 221 nm thick CYTOP, as this device remained stable without breakdown up to the maximum applied electric field of ≈9 MV cm$^{-1}$), along with the extracted results from reported studies on various transferred insulators[14-16,19,20] evaluated in MIM configurations. The comparison reveals that the transferred CYTOP films exhibit exceptional insulating characteristics, sustaining among the highest breakdown fields reported for transferred dielectrics. The high breakdown field of CYTOP makes it a highly suitable gate dielectric for transistors, ensuring low leakage and stable device operation under high-voltage conditions. Furthermore, its low dielectric constant enables reduced parasitic capacitance, which is advantageous for high-frequency transistor operation and also for inter-metal dielectric applications, where it can minimize capacitance between densely packed interconnects. The reduction in capacitance mitigates resistance-capacitance delays, suppresses signal cross-talk, and lowers power dissipation, thereby enabling high-frequency and power-efficient operation.[34-36]

### 3. H-terminated Diamond FETs with Transferred CYTOP Films as Gate Insulator

To validate the effectiveness of our lamination-based approach and to demonstrate the formation of a high-quality dielectric/substrate interface, H-terminated diamond FETs were fabricated by employing CYTOP as the gate insulator. Owing to its ultra-wide bandgap, high thermal conductivity, high carrier mobility and large breakdown field,[37] diamond has emerged as a promising candidate to fabricate FETs for advanced electronic applications.[38-45] The H-terminated diamond exhibits high valence band maximum, which facilitates the generation of two-dimensional hole gas (2DHG) at the surface. This 2DHG can be generated either by applying a gate voltage[39,43,44] or, more commonly, through surface transfer doping by exposing the H-terminated diamond surface to ambient air,[46] gases such as $NO_2$ and $O_3$,[47] or deposition of high-electron-affinity oxides like $MoO_3$ and $V_2O_5$.[48] The surface transfer doping has widely been used to fabricate *p*-channel FETs, offering practical alternatives, given the high activation energies associated with conventional dopants in diamond.[46] However, the surface transfer doping introduces ionized acceptors or fixed negative charges on the diamond surface, which scatter carriers and limit device performance. To address these limitations, H-terminated diamond FETs were fabricated without surface transfer doping by processing the devices in an inert environment using *h*-BN as a gate insulator, which showed high channel mobility.[39,44] Here, we use transferred CYTOP film as a gate insulator while avoiding surface transfer doping by processing the lamination on H-terminated diamond in inert environment. This offers a simple and large-area-compatible approach to fabricating high-performance H-terminated diamond FETs.

In the present work, (100)-oriented single-crystalline undoped diamond substrates grown by CVD were used. To define the active channel and non-conductive regions, nitrogen ion implantation was performed on diamond substrates by masking the desired regions for active channel with patterned photoresist, as shown in Figure S3(a). Nitrogen-implanted regions in diamond maintain high resistivity even after H-termination and subsequent air exposure.[49,50] Following implantation and photoresist mask removal, Ti/Pt (5 nm/5 nm) electrodes were patterned on the defined active channel regions (Figure S3(b)). The samples were then transferred to microwave-plasma assisted CVD (MPCVD) system and annealed in $H_2$ environment at a stage temperature of 650 °C for 35 minutes, followed by hydrogen plasma exposure at 570 °C for 15 minutes, enabling titanium-carbide (TiC) ohmic contact formation and H-terminated diamond surface. Note that during the hydrogen plasma treatment, the actual diamond surface temperature reached approximately 700 - 750 °C, as measured by a pyrometer. The H-terminated diamond was subsequently transferred to an Ar-filled glove box using a vacuum suitcase, where CYTOP lamination was carried out without air exposure to avoid surface transfer doping. The detailed procedures for FET fabrication are provided in the *Experimental Section*.

Figure 4 presents the room temperature electrical characteristics of a H-terminated diamond FET with a 90 nm thick CYTOP gate insulator (see also Supporting Information Section 4 and Figure S5 for the characteristics of second H-terminated diamond FET). The schematic of H-terminated diamond FET using CYTOP gate insulator is illustrated in Figure 4(a), while the optical microscope image of FET is shown in Figure 4(b). The output curves of the FET are shown in Figure 4(c), in which the drain current ($I_{DS}$) was measured by sweeping the drain voltage ($V_{DS}$) from 0 to −30 V and then in the reverse direction at various gate voltages ($V_{GS}$), which were varied from +2 V to −30 V in 2 V steps. These results show typical *p*-type FET behavior and negligible hysteresis. The output curves demonstrate clear current modulation across both the linear and saturation regions, confirming effective field-effect control of 2DHG at the diamond surface. The geometrical parameters of Hall-bar configuration, including the channel length ($L_G$), channel width ($W_G$), and the separation between voltage probes ($L_p$), were determined from optical microscope images, following the procedure described in our previous work.[39] For this particular device, $L_G$ was estimated to be 117.5 μm, $W_G$ was 8.4 μm, and $L_p$ was 29.8 μm. As shown in Figure 4(c), the drain current was normalized by dividing the values with $W_G$. The maximum drain current was observed to be −23 mA mm$^{-1}$ at $V_{DS} = V_{GS} = -30$ V, which is considerably high given the relatively long gate length of 117.5 μm.

Figure 4(d) presents the transfer characteristic of the FET measured in two-probe configuration, where the drain current is plotted on a logarithmic scale as a function of gate voltage at $V_{DS} = -30$ V, from which an on/off current ratio of $5.7 \times 10^6$ was extracted. The normalized hysteresis width, obtained from Figure 4(d) in the gate voltage range of 0.5 to −1 V, was found to be <5 mV (MV cm$^{-1}$)$^{-1}$, which is comparable to the lowest values reported for 2D semiconductors employing the transferred gate insulators.[14,18] The hysteresis width was normalized by dividing the width by maximum applied field (*i.e.* maximum $V_{GS}$ divided by the CYTOP thickness). The subthreshold swing (SS) from Figure 4(d) was determined to be 220 mV dec$^{-1}$, corresponding to a low interface trap density of $\leq 3 \times 10^{11}$ cm$^{-2}$ eV$^{-1}$, which was obtained following the procedure described in Supporting Information. The interface trap density in the present case is comparable to or lower than most reported values for H-terminated diamond FETs,[39,45,51-54] which indicates that a clean and high-quality interface is achieved through our freestanding CYTOP transfer method and fabrication process that avoids ambient air exposure of the H-terminated diamond surface. The reduced trap density is attributed to the non-destructive lamination of the CYTOP film, which substantially avoids any distortion on the H-terminated diamond surface in contrast to direct dielectric deposition, as well as to the inherent absence of dangling bonds in CYTOP. The sheet conductance shown in Figure 4(e) was measured using a four-probe configuration by sweeping $V_{GS}$ from +5 V to −30 V and then in the reverse direction, while applying $V_{DS}$ of ±100 mV. The maximum sheet conductance was $2.1 \times 10^{-4}$ Ω$^{-1}$, corresponding to a sheet resistance of 4.8 kΩ. By extrapolating the linear region of sheet conductance in the gate voltage range of −5 V to −15 V, the threshold voltage ($V_{th}$) was estimated to be −2.3 V. Figure 4(f) presents the calculated field-effect mobility ($\mu_{FE} = \frac{t_{CYTOP}}{\epsilon_{CYTOP}}\left|\frac{\partial\sigma}{\partial V_{GS}}\right|$) and effective mobility ($\mu_{eff} = \frac{t_{CYTOP}}{\epsilon_{CYTOP}}\frac{\sigma}{|V_{GS} - V_{th}|}$) as functions of gate voltage by following the reported procedure.[39] Here, $t_{CYTOP}$ and $\epsilon_{CYTOP}$ are the thickness and dielectric constant of CYTOP films, respectively, and σ is the sheet conductance. The maximum effective mobility was calculated to be ≈400 cm$^2$V$^{-1}$s$^{-1}$. A maximum hole density reached to $3.2 \times 10^{12}$ cm$^{-2}$ at $V_{GS} = -30$ V. The estimated mobility in CYTOP-based diamond FETs is significantly high and falls within the upper range of reported values using various gate insulators.[39-45,51-54] The stability of the FET after long-term storage was also evaluated, demonstrating nearly identical output and transfer characteristics, as discussed in the Supporting Information. Figure 4(g) shows the gate leakage current density as a function of electric field, where the leakage current was measured by shorting the source and drain electrodes and sweeping $V_{GS}$ from +4 V to −30 V and then in reverse direction. The leakage current was normalized to the active channel area under the gate electrode (8328 μm$^2$), and the gate voltage was converted to electric field by dividing by the CYTOP film thickness. The resulting current density remains below $5 \times 10^{-7}$ A cm$^{-2}$ at an electric field of 3 MV cm$^{-1}$. The electric field value is found to be larger than or comparable to those reported for diamond FETs using other gate dielectrics.[51-58]

To further investigate the interface quality of CYTOP/H-terminated diamond as well as the dielectric properties of the transferred CYTOP film, *C-V* measurements of the diamond FET were performed at different frequencies in the range of 10 kHz to 1 MHz in $N_2$-filled glove box, by sweeping $V_{GS}$ in the range of +4.0 V and −4.1 V in both forward and reverse directions. The *C-V* curves were obtained from a three-terminal FET device fabricated on the same substrate, as shown in Figure 4(b). The channel width and length, in this case, are 41.9 μm and 8.4 μm, respectively. To minimize the influence of parasitic capacitances on the estimation of the insulator capacitance, the source and drain electrodes were electrically shorted during the measurements. Figure 4(h) presents the *C-V* characteristics obtained using an AC signal of 50 mV, measured across various gate voltages and frequencies, in which the capacitance was normalized to the channel area. Initially, the *C-V* measurements were performed between the gate and the shorted electrodes at frequencies ranging from 10 kHz to 1 MHz with the $V_{GS}$ set to 0 V. Similarly, a series of measurements over a gate voltage sweep from +4.0 V to −4.1 V with 0.3 V step. Figure 4(h) displays OFF-state and ON-state regions. For the case of 1 MHz frequency, the measured capacitance in ON-state regime was found to be 0.126 μFcm$^{-2}$ and remained relatively stable up to $V_{GS}$ = 0 V. As the device transitioned toward the OFF-state, the capacitance began to drop, reaching a steady value of 0.106 μFcm$^{-2}$ at $V_{GS}$ = 1 V and remains unchanged thereafter. The OFF-state capacitance arises due to the overlap of source-drain and gate electrodes, which also includes the parasitic contribution mainly from probe needles. Conversely, in the ON-state, the total measured capacitance includes both the OFF-state contribution and the capacitance of the gate dielectric corresponding to the conductive channel. Therefore, the capacitance of CYTOP can be estimated by subtracting the OFF-state value from the ON-state value. Hence, the capacitance associated with the CYTOP layer under channel region is found to be 0.020 $\mu Fcm^{-2}$, which corresponds to a dielectric constant of 2.0 $\varepsilon_0$ (where $\varepsilon_0$ is the vacuum permittivity), given the CYTOP thickness of 90 nm. Notably, the dielectric constant remained consistent across different measurement frequencies and the value aligns well with reported values for CYTOP.[21] The *C-V* measurements show nearly overlapping curves for the forward and reverse sweeps, which corroborate well with the transfer curve (Figure 4(d)) and further confirm a substantially low interface trap density at the CYTOP/H-terminated diamond interface.

**4. Discussion**

The transfer and lamination method demonstrated in this work has broad applicability for the integration of fluoropolymer dielectric onto the substrates that are considered to be incompatible with conventional solution-based methods like spin coating or dip coating. For example, H-terminated diamond surface exhibits high water contact angles and strong liquid repellency,[49,59] which leads to poor wettability and nonuniform film formation when polymer solutions are applied. By following our transfer method, CYTOP films were laminated without affecting the H-termination by exposing the surface to ambient air or any solvents. Hence, this approach enables the integration of preformed and thermally cured dielectric films onto solvent or heat sensitive substrates without distorting the surface properties by any solvents and heat treatments. This method further holds significant promise for encapsulation and passivation in flexible electronics, in which solvent resistance and low temperature process are essential. Additionally, the transfer of CYTOP films may also be suitable for protecting moisture sensitive devices due to its low water vapor permeability, as well as have potential for optoelectronics and light emitting devices owing to high optical transparency of CYTOP. Furthermore, the lamination of CYTOP films onto diamond substrates may open new possibilities in quantum technologies, particularly in devices employing nitrogen vacancy (NV) centers in diamond for sensing or quantum information processing. Owing to the high optical transparency of CYTOP and non-destructive integration, it is expected to preserve the optical quality and spin coherence vital for NV center functionality.[60]

From the electronic device perspective, the present study employed type-S CYTOP, which has been shown to exhibit an electron affinity of approximately 3.6 eV based on low-energy inverse photoelectron spectroscopy (LEIPS).[61] Accordingly, the lowest unoccupied molecular orbital (LUMO) level of type-S CYTOP is estimated to lie near −3.6 eV with respect to the vacuum level, which is energetically higher than the valence band maximum (VBM) of H-terminated diamond (−4.2 eV).[62] In addition, CYTOP exhibits optical absorption in the ultraviolet region below 200 nm,[21] suggesting an energy gap between LUMO and highest occupied molecular orbital (HOMO) levels exceeding 6 eV. This indicates that the HOMO level of CYTOP is located significantly deeper than VBM of H-terminated diamond, thereby enabling stable p-channel FET operation. Under ideal conditions where the LUMO level of type-S CYTOP lies above the VBM of H-terminated diamond, electron transfer from the valence band of H-terminated diamond to the LUMO of S-type CYTOP is energetically unfavorable. Nevertheless, lamination of the CYTOP film onto H-terminated diamond induced a slight surface conductivity, as shown in the Supporting Information (Figure S4), although the conductivity remained significantly lower than that reported for other gate dielectrics such as $Al_2O_3$.[63] The emergence of the surface conductivity may arise from charge trapping sites within the CYTOP film, which can accept electrons from the valence band of H-terminated diamond and generate holes at the diamond surface, similar to the surface transfer doping mechanism reported for H-terminated diamond with deposited gate dielectrics.[46] Such trapping sites in the CYTOP could originate from the distortion of polymer chains due to local strain,[64] or residual solvent and adsorbate species, although their exact origin and energetic distribution remain unclear at present. Further experimental and theoretical investigations are therefore required to clarify the electronic interactions responsible for the observed conductivity following CYTOP lamination. Other variants, such as type-M and type-A CYTOP, have been reported to exhibit larger electron affinity values,[62] positioning their LUMO level comparable to or below the VBM of H-terminated diamond. Therefore, direct lamination of these types of CYTOP on H-terminated diamond could lead to a more pronounced charge transfer even in the absence of charge trapping sites. This would enhance the generation of holes, but the increased amount of transferred charge in the CYTOP would act as a scattering source, limiting the hole mobility. An interesting future strategy for generating a high hole density while maintaining high mobility could involve a modulation doping scheme[65,66] by stacking different types of CYTOP layers, where a low electron affinity CYTOP is first laminated on H-terminated diamond followed by a higher electron affinity CYTOP layer. Such stacking approach enables the creation of functional heterostructures with well-defined layers, which are challenging to achieve by conventional solution-based methods due to intermixing of polymer molecules, and therefore offers an additional merit of the film transfer method.

## 5. Conclusions

In summary, we demonstrated a simple and effective approach for the transfer of freestanding low-$\kappa$ amorphous fluoropolymer films, enabling their integration onto various substrates, including low-adhesive and sensitive substrates like H-terminated diamond. This process allows the transfer of CYTOP dielectric films in ambient air and inert atmospheres, without compromising the integrity of dielectrics or surface quality of the target substrate. The MIM structures fabricated by the integration of freestanding CYTOP films have shown high dielectric breakdown fields, indicating high quality and robustness of the insulator. Notably, *p*-channel FETs fabricated on H-terminated diamond using CYTOP gate insulators exhibit excellent electrical performance, including high hole mobility, negligible hysteresis, and substantially low interface trap density, validating the high-quality CYTOP/H-terminated diamond interface. Overall, these results validate the efficacy of the proposed transfer method and highlight the excellent dielectric properties of CYTOP films. This work establishes a promising large-area-compatible integration strategy, offering a new pathway for creating high-quality interfaces on sensitive surfaces, holding significant potential for utilizing CYTOP in a wide range of advanced electronic and optoelectronic applications.

## 6. Experimental Section

**6.1. Preparation and Transfer Method of Freestanding CYTOP Films:** 3 wt% fluoropolymer solution was prepared by dissolving type-S CYTOP (CTX-809SP2)[21] in CT-Solv.180 solvent (both from Asahi Glass Co., Ltd., Japan). To ensure complete dissolution and homogenization, the solution was magnetically stirred for 2 hours. Parallelly, a 3 wt% PAA solution was prepared by dissolving PAA powder (average molecular weight = 5000 g $mol^{-1}$, Wako Pure Chemicals) in DI water. Single-side polished Si substrates with the approximate size of 2 cm × 2 cm were subjected to a sequential ultrasonic cleaning in DI water, isopropanol (IPA), acetone, IPA, and finally DI water to remove organic and particulate surface contaminants. The substrates were then treated with oxygen plasma (20 sccm $O_2$ flow, 10 Pa chamber pressure, 250 W RF power, 3 minutes at room temperature) to further eliminate organic residues and enhance surface energy for better film adhesion. After plasma treatment, Si substrates were first coated with PAA solution by spin-coating at 3000 rpm for 60 s, followed by baking at 155 °C for 2 minutes to evaporate residual water and enhance film integrity. Subsequently, the CYTOP solution was spin-coated onto the PAA layer under identical parameters and underwent stepwise thermal annealing: 50 °C for 10 min (solvent removal), 80 °C for 30 min (film densification) and 130 °C for 10 min (curing and mechanical reinforcement). This resulted in uniform and substantially smooth CYTOP films with approximate thickness of 90 nm, suitable for exfoliation from substrate by dissolving PAA sacrificial layer. The thickness of CYTOP films can be controlled by varying the speed of spin coating (rotation speed) or the concentration of CYTOP solution.[21] To enable exfoliation, high-temperature silicon-adhesive double-sided Kapton tape (KTP36-3/4SD, TECSAM, Japan; maximum operating temperature 260 °C) was cut to ≈1.9 cm × 2.5 cm with a central ≈6 mm × 6 mm aperture, forming a transfer window, and affixed onto the CYTOP film. Although the Kapton tape is double-sided, only one side was used leaving the protective layer on the opposite side intact. The resulting Kapton/CYTOP/PAA/Si stack was floated on a DI water bath at 50 °C to accelerate dissolution of the underlying PAA layer. After complete PAA dissolution, the Si substrate was detached, leaving the Kapton-supported CYTOP film floating on the water surface. For improved alignment and handling, the freestanding CYTOP film was mounted onto a 2 mm-thick copper frame (2.6 cm × 8.6 cm) containing a 4.5 mm circular aperture near one edge. For the transfer or lamination of freestanding CYTOP films onto target substrate, an x-y-z direction movable stage to hold the copper frame was used, while the target substrate was mounted on an adjacent stage (equipped with x-y translation and rotation adjustment) with vacuum chuck for stable lamination under an optical microscope (Figure 2).

**6.2. Characterization of CYTOP Films:** The CYTOP film thickness was measured using Thickness Monitor FE-3000 (Otsuka Electronics), with refractive index values of CYTOP[21] calibrated for the relevant spectral range. A thickness variation of ±5 nm was observed across samples prepared from separate CYTOP solutions with identical spin-coating parameters, confirming the process reproducibility. Furthermore, the thickness measurements conducted at multiple locations on the same sample showed a negligible variation of ±0.6 nm, indicating the overall uniformity of CYTOP films. The thickness of PAA layer was measured using Bruker's DektakXT profiler, which was estimated to be 33 ± 4 nm. AFM was performed using AFM5100N (Hitachi High-Tech) to evaluate surface morphology and roughness. MIM structures were fabricated using CYTOP as the dielectric film on polished sapphire substrates with approximate size of 2 mm × 2 mm × 0.5 mm. The substrates were cleaned by dipping in a base Piranha solution consisting of DI water, ammonium hydroxide and hydrogen-peroxide at 80 °C, followed by the ultrasonic cleaning in organic solvents, as mentioned above. Bottom electrodes were patterned with the widths of 5 μm and 10 μm on the same substrate (inset of Figure 3(b)) using maskless laser lithography (DL-1000, Nano System Solutions) and metal (Ti/Pt, 5 nm/5 nm) deposition via e-beam evaporation. CYTOP film was laminated onto the patterned substrate by following the transfer method described above and the top electrodes (Al/Au, 50 nm/100 nm) were deposited through a shadow mask using a custom-built thermal evaporator to complete the MIM devices. To determine the current density, the overlapping area of CYTOP films between bottom and top electrodes was estimated from optical microscope images by plotting brightness versus distance, and fitting the resulting profile using an error function.[39]

**6.3. H-terminated Diamond Field-Effect Transistors:** CVD grown standard grade IIa-type (100) single-crystalline diamond substrates, procured from Element Six, were employed to fabricate FETs in this study. The diamond substrates were polished to minimize surface roughness and diced into pieces of approximately 2.2 mm × 2.2 mm × 0.5 mm. The substrates were then subjected to sequential ultrasonic cleaning in DI water, IPA, acetone, IPA and DI water to remove organic and particulate contaminants. To isolate device regions, nitrogen ion implantation ($^{15}N$, 10 keV, $1 \times 10^{14}$ ions $cm^{-2}$) was performed at room temperature by partially covering diamond with photoresist to protect the channel region. After the implantation, photoresist mask was removed and substrates were cleaned using a hot acid mixture ($HNO_3$:$H_2SO_4$ = 1:3) at 200 °C for 30 minutes, followed by a final organic cleaning step. The electrodes were patterned for electrical measurement using laser lithography and subsequent deposition of Ti/Pt (5 nm/5 nm) using e-beam evaporation. After formation of Ti/Pt electrodes, the sample was transferred to MPCVD (Seki Technotron, AX-5000) system and annealed in $H_2$ gas at a temperature of 650 °C for 35 min. After annealing, the sample was exposed to hydrogen plasma at 570 °C for 15 minutes with a hydrogen flow rate of 500 sccm and chamber pressure of 27 Torr. These steps facilitated the formation of TiC-based ohmic contacts and hydrogen termination of the diamond surface, as reported previously.[39] After H-termination, diamond substrate was transferred to an Ar-filled glove box using a vacuum suitcase, without exposing diamond surface to the ambient air. The vacuum suitcase is equipped with a non-evaporable getter (NEG) pump, which maintains pressure below $1 \times 10^{-7}$ Torr. The high-purity of Ar inside the glove box was maintained by circulating the Ar gas through a purifier, which maintained oxygen and water contents below 0.5 ppm and 2 ppm, respectively. In order to remove any atmospheric adsorbates, the freestanding CYTOP film supported by copper frame was baked inside Ar-filled glove box at temperatures of 80 °C, 100 °C and 120 °C for 3 minutes each to remove atmospheric adsorbates before transferring the H-terminated diamond from the vacuum suitcase. No change in the purity of Ar was observed inside glove box during the CYTOP film baking, suggesting minimum adsorbates on the film. After this step, the H-terminated diamond was immediately transferred to the Ar-filled glove box and CYTOP lamination was performed, following the methodology described above. The lamination was completed within 5 hours after transferring the diamond from the MPCVD system to the Ar-filled glove box. Following lamination, a metal shadow mask was fixed to the sample inside the glove box. The sample was then transferred via a vacuum suitcase to a thermal evaporator, where gate electrode (Al/Au, 50 nm/100 nm) was deposited. Finally, the completed FET device was transferred via vacuum suitcase to an $N_2$-filled glove box for electrical characterization.

**6.4. Electrical Measurement Setup:** The electrical measurements were performed inside the $N_2$-filled glove box using probe station by directly touching the gold-alloy probe needles with a tip diameter of 20 μm to the electrode pads on the substrate. The output and transfer characteristics were measured on the gated Hall-bar sample (Figure 4(b)), in two-terminal and four-probe configurations, respectively. The gate bias was applied using a source measure unit (Keysight Technologies, B2901A), while the drain voltage was provided by a function generator (Agilent Technologies, 33220A). A voltage amplifier (DL Instruments, 1201) and current preamplifier (DL Instruments, 1211) were used to measure longitudinal voltage and drain current, respectively. The sheet resistance was obtained by applying a voltage of ±100 mV between source and drain electrodes. Sheet resistance ($\rho$) was calculated using the formula $\rho = (V_{p+} - V_{p-})/(I_{D+} - I_{D-}) \times (W_G/L_p)$, where $V_{p+}$ and $V_{p-}$ are the measured voltages across the longitudinal probes for positive and negative currents, respectively. The sheet conductance ($\sigma$) was calculated as the reciprocal of $\rho$. Output characteristics were obtained using the source measure unit for gate biasing, a source meter (Keithley instruments, 2410) for drain voltage and the current preamplifier (DL Instruments, 1211) to measure the drain current. An electrometer (Keithley instruments, 6514) was used to measure drain current to calculate the ON/OFF ratio. To measure the gate leakage current, the source and drain electrodes were shorted together and the current was measured while applying voltage to the gate electrode. The capacitance-voltage characteristics were measured using a precision impedance analyzer (Agilent Technologies, 4294A). To minimize the effects of parasitic capacitance, coaxial probe needles were employed, in which only a small portion of the gold-alloy needle was exposed.

**Competing interests**

The authors declare no competing interests.

**Acknowledgements**

We thank T. Kageura and S. Onoda for their helpful discussions and M. Imura, T. Uchihashi and T. Ando for their support. We also thank N. Mitsunaga and H. Sakai for their preliminary work on the transfer of CYTOP films, and Y. Shimono for performing AFM measurements. This study was financially supported by the JSPS KAKENHI (Grant Nos. JP22H01962, JP23K23230), the NEDO Uncharted Territory Challenge 2050 (Project No. JPNP14004), and the Advanced Research Infrastructure for Materials and Nanotechnology in Japan (ARIM) (Proposal Nos. JPMXP1224NM5227, JPMXP1225NM5207).

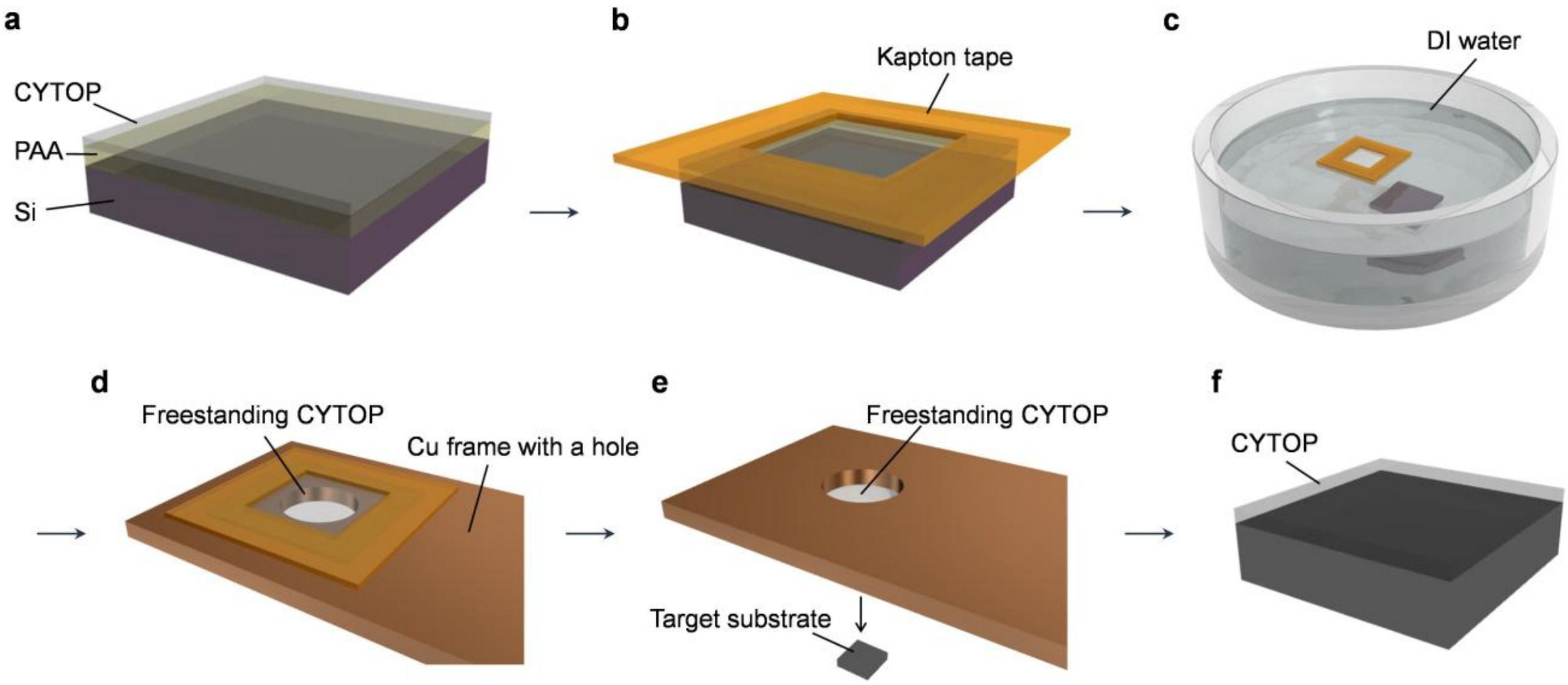

**Figure 1.** Schematic illustration of the CYTOP film transfer process. (a) Spin coated CYTOP and PAA layers on Si substrate. (b) Attachment of a Kapton tape on CYTOP surface for mechanical support. (c) Release of Kapton-supported CYTOP film by dissolving the PAA layer in water. (d) Mounting of the Kapton-supported CYTOP film onto a copper frame with an aperture. (e) Lamination of the freestanding CYTOP film onto the target substrate. (f) Target substrate laminated with CYTOP film.

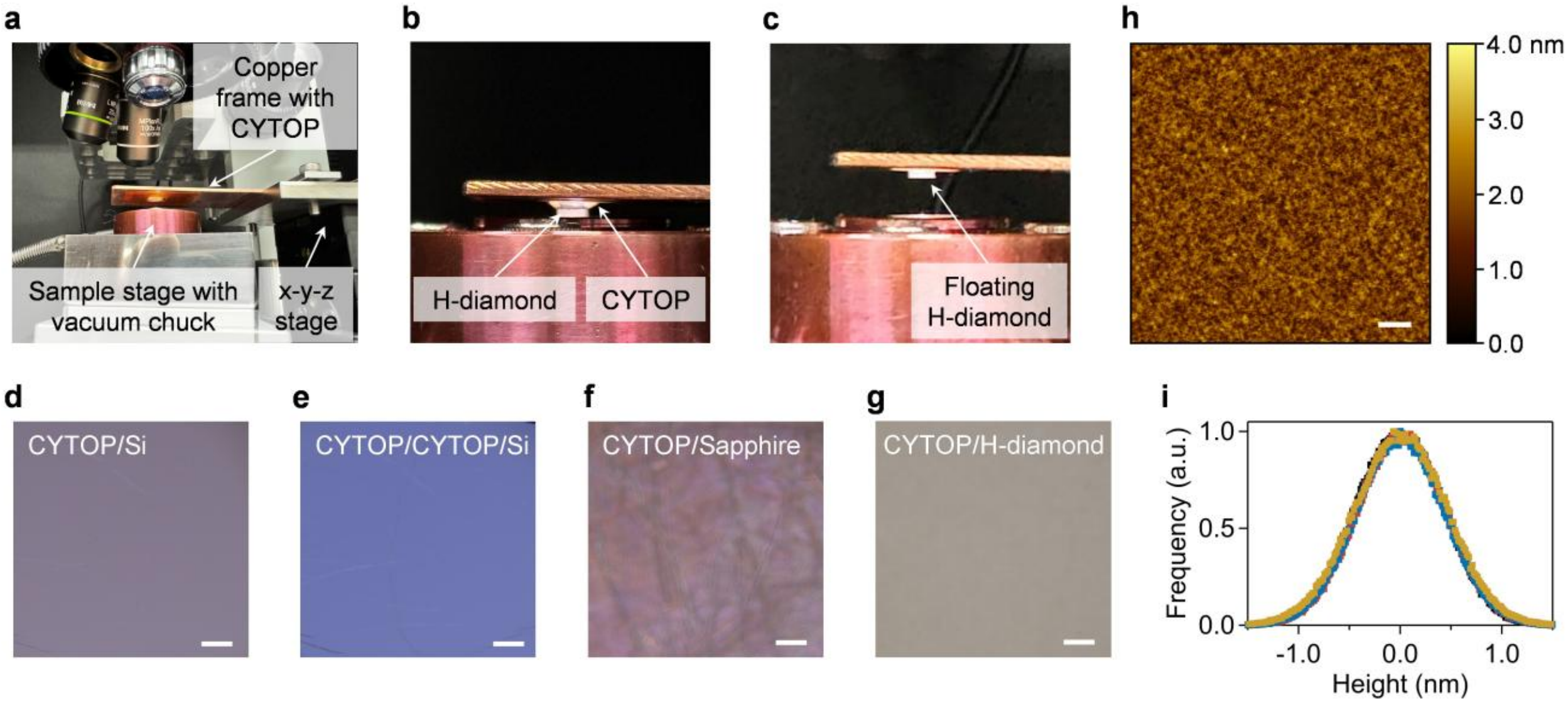

**Figure 2.** The transfer of freestanding CYTOP films. (a) Sample stage with vacuum chuck along with an adjacent x-y-z movable stage holding the copper frame containing the freestanding CYTOP film. (b) Laminated CYTOP film on H-terminated diamond in ambient air, showing CYTOP bending at the diamond periphery due to slight upward movement of the copper frame while the diamond remains fixed on the stage. (c) H-terminated diamond floating after release from the sample stage and further upward movement of the copper frame. CYTOP film laminated on (d) Si, (e) CYTOP/Si, (f) sapphire, and (g) H-terminated diamond. (The dark lines observed in (f) do not correspond to features of the CYTOP films, but originate from the underlying sample stage, made visible by the high transparency of the sample, and further enhanced by slight focus deviations.) Scale bars in (d)-(g): 200 µm. (h) AFM image of a spin-coated CYTOP film on PAA/Si (Scale bar, 2 µm). (i) The height histogram obtained for four different samples.

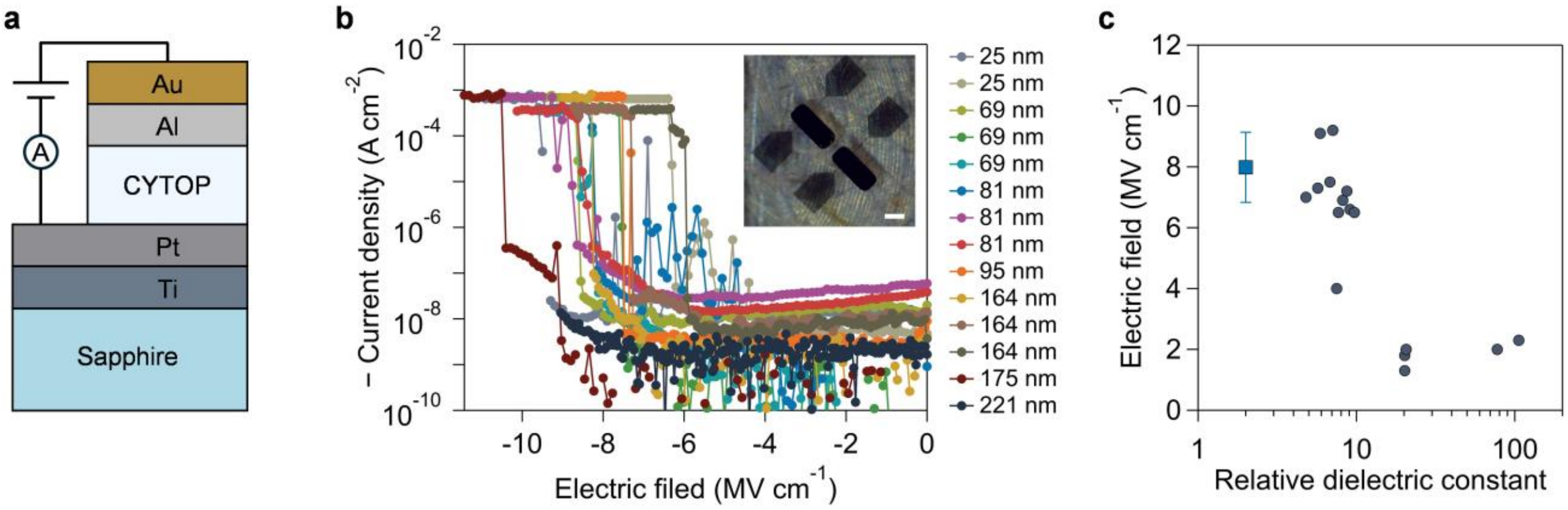

**Figure 3.** Metal-insulator-metal (MIM) structure. (a) Schematic illustration of the MIM structures. (b) Leakage current density as a function of electric field for MIM structures fabricated with transferred, freestanding CYTOP dielectric films of varying thicknesses. The data for the 175 nm device corresponds to a two-layer stack of CYTOP films. The measured current was converted to current density using the overlapping area of the CYTOP film between the top and bottom electrodes, while the voltage was converted to electric field by normalizing it to the CYTOP film thickness. Negative sign of the electric field corresponds to negative voltage applied to the top electrode (Al/Au) with respect to the bottom electrode (Ti/Pt). The current density after the breakdown corresponds to the source meter compliance limit. The inset in (b) shows an optical microscope image of a representative MIM structure fabricated on sapphire with CYTOP as the dielectric. Scale bar: 200 μm. (c) Comparison of the electric field at a current density of $10^{-5}$ A cm$^{-2}$ for different transferred gate insulators versus relative dielectric constant. Blue square represents the average value from 13 devices (excluding the device with 221 nm thick CYTOP) measured in this work, where error bar corresponds to the standard deviation, while grey circles denote reported data from the literature.[14-16,19,20]

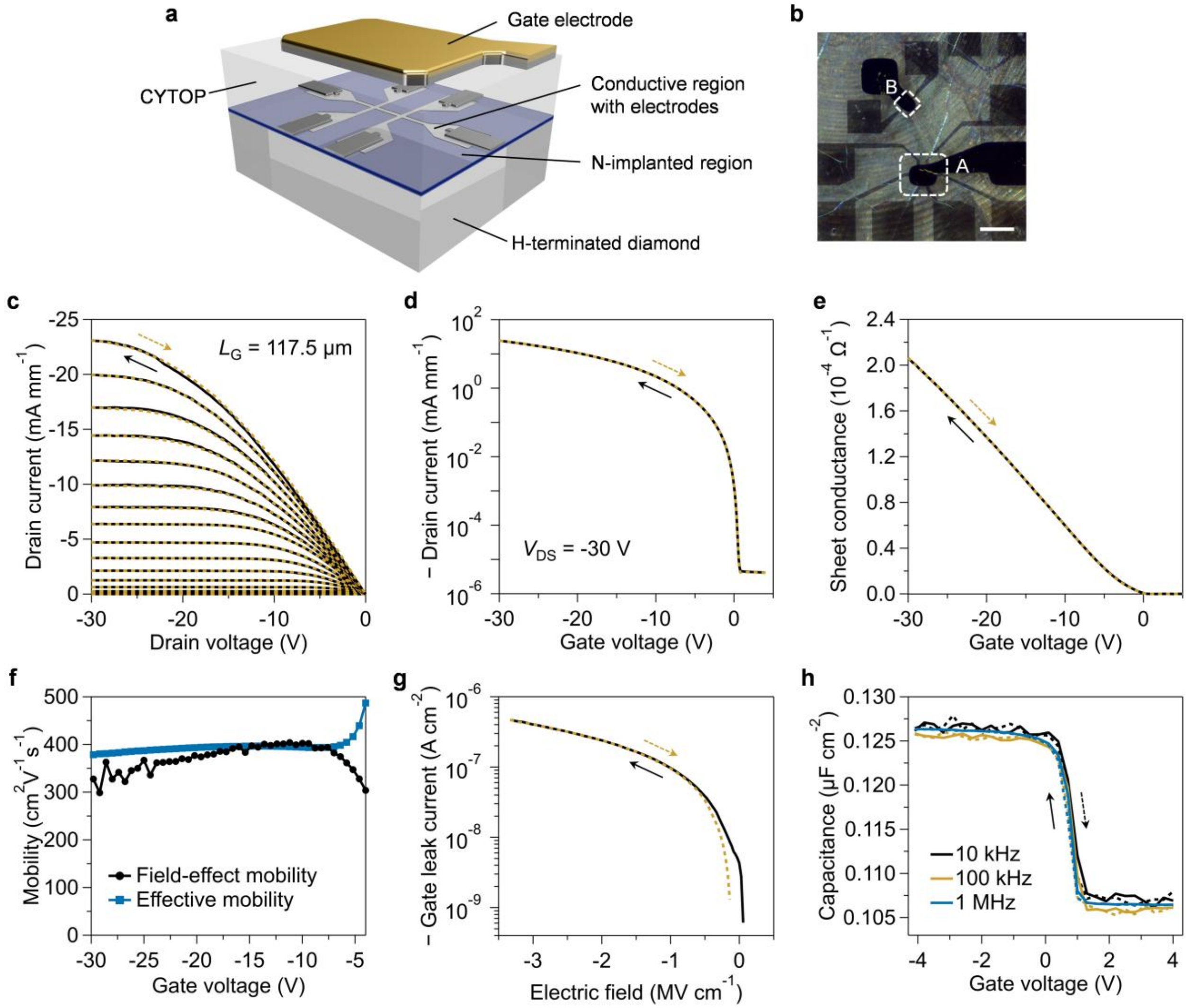

**Figure 4.** Electrical characteristics of the diamond FET with a CYTOP gate insulator at room temperature. (a) Schematic illustration of the CYTOP-gated diamond FET with Hall-bar geometry on diamond. (b) Optical microscope image of the FET. Scale bar, 200 μm. (c) Output characteristics showing drain current versus drain voltage for gate voltages varied from +2 to −30 V in 2 V steps. (d) Drain current versus gate voltage at a drain voltage of −30 V, with the gate voltage swept from +4 to −30 V and back. (e) Gate voltage dependence of sheet conductance in the linear region obtained from four-probe measurements, where the gate voltage was swept from +5 to −30 V and back. (f) Effective and field-effect mobilities as a function of gate voltage. (g) Gate leakage current density as a function of electric field. The electrical characteristics shown here were obtained for the device enclosed by dashed line A in (b), which has a gated Hall-bar structure as shown in (a). In (c), (d), (e), and (g), solid black and dashed brown lines correspond to forward and backward sweeps, respectively. (h) Capacitance-voltage (*C*-*V*) curves obtained from the three-terminal FET device enclosed by dashed line B in (b) at different frequencies, where solid and dashed lines represent the measurements from +4.0 to −4.1 V and from −4.1 to +4.0 V, respectively.

# Supporting Information

# Transfer of Freestanding Fluoropolymer Films for Advanced Semiconductor Devices

Mohammad Monish[1,*], Koki Hino[1,2], Yosuke Sasama[3], Masato Urakami[4], Takehiro Ota[4], Kenji Sakamoto[5]
Kenichiro Takakura[4], Yamaguchi Takahide[1,2,*]

[1]Research Center for Materials Nanoarchitectonics, National Institute for Materials Science, Tsukuba, Ibaraki 305-0044, Japan
[2]University of Tsukuba, Tsukuba, Ibaraki 305-8571, Japan
[3]International Center for Young Scientists, National Institute for Materials Science, Tsukuba, Ibaraki 305-0044, Japan
[4]National Institute of Technology, Kumamoto College, Koshi, Kumamoto 861-1102, Japan
[5]Research Center for Macromolecules and Biomaterials, National Institute for Materials Science, Tsukuba, Ibaraki 305-0047, Japan

## 1. Freestanding CYTOP Film Lamination

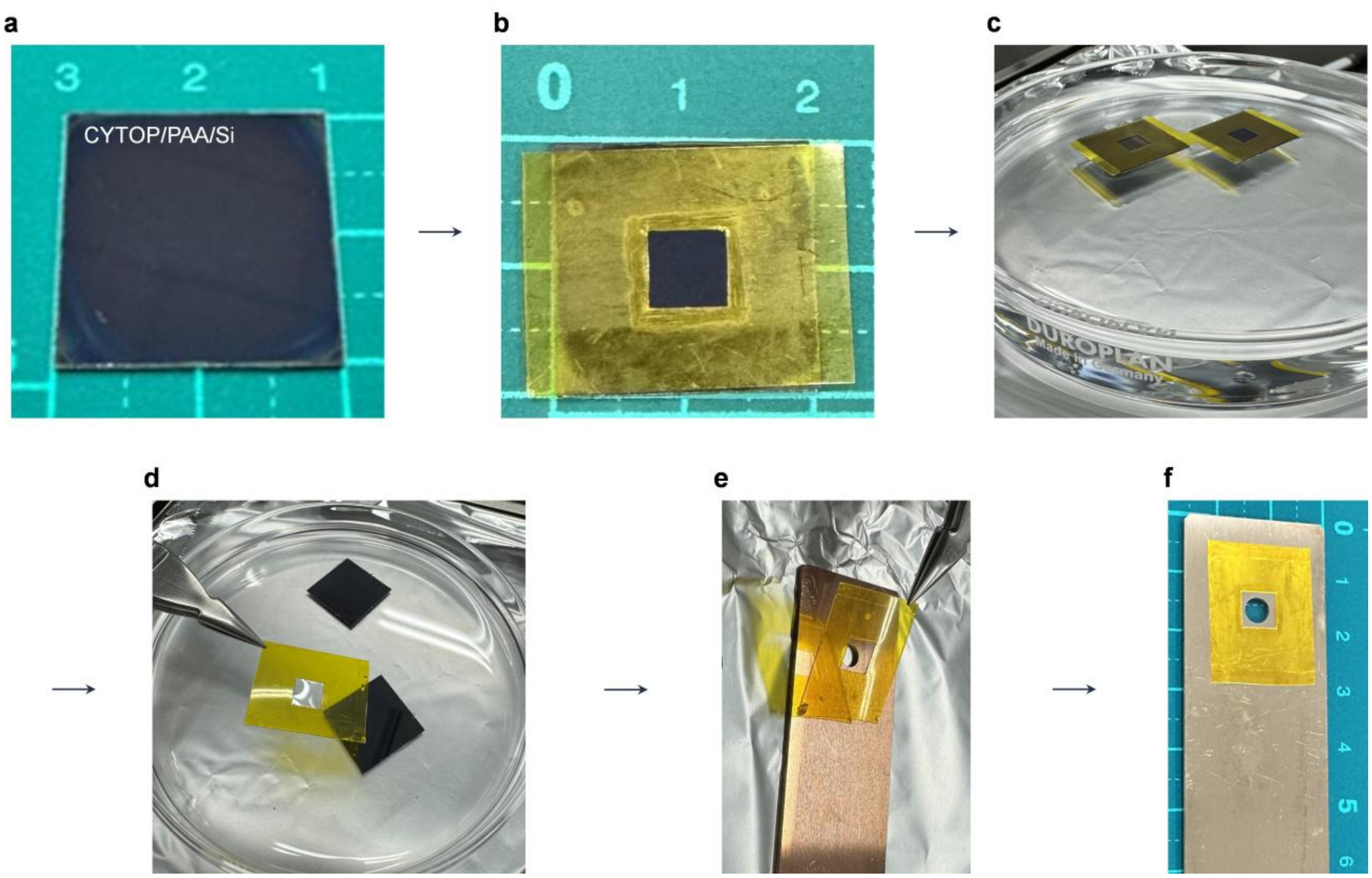

**Figure S1.** Photographs showing preparation of freestanding CYTOP film. (a) Spin-coated CYTOP and PAA layers on a Si substrate. (b) Attachment of Kapton tape onto the CYTOP/PAA/Si stack. (c) Kapton tape-supported assembly floated on DI water. (d) Kapton tape-supported freestanding CYTOP film. (e) Mounting CYTOP onto copper frame with a circular aperture. (f) Final CYTOP film assembly.

As described in the main manuscript, a polyacrylic acid (PAA) sacrificial layer was first spin-coated on Si substrates, followed by spin coating of CYTOP film. Figure S1(a) shows the CYTOP/PAA/Si sample (2 cm × 2 cm) after baking. A Kapton tape with a central cut was attached on top of the CYTOP film to provide wrinkle-free support during transfer (Figure S1(b)). The sample was then floated in deionized (DI) water (Figure S1(c)), dissolving the PAA layer and releasing the freestanding CYTOP film attached to the Kapton tape (Figure S1(d)). The film was subsequently mounted onto a copper frame with a circular aperture (Figures S1(e) and S1(f)) to facilitate precise lamination onto various substrates.

## 2. Atomic Force Microscopy Analysis

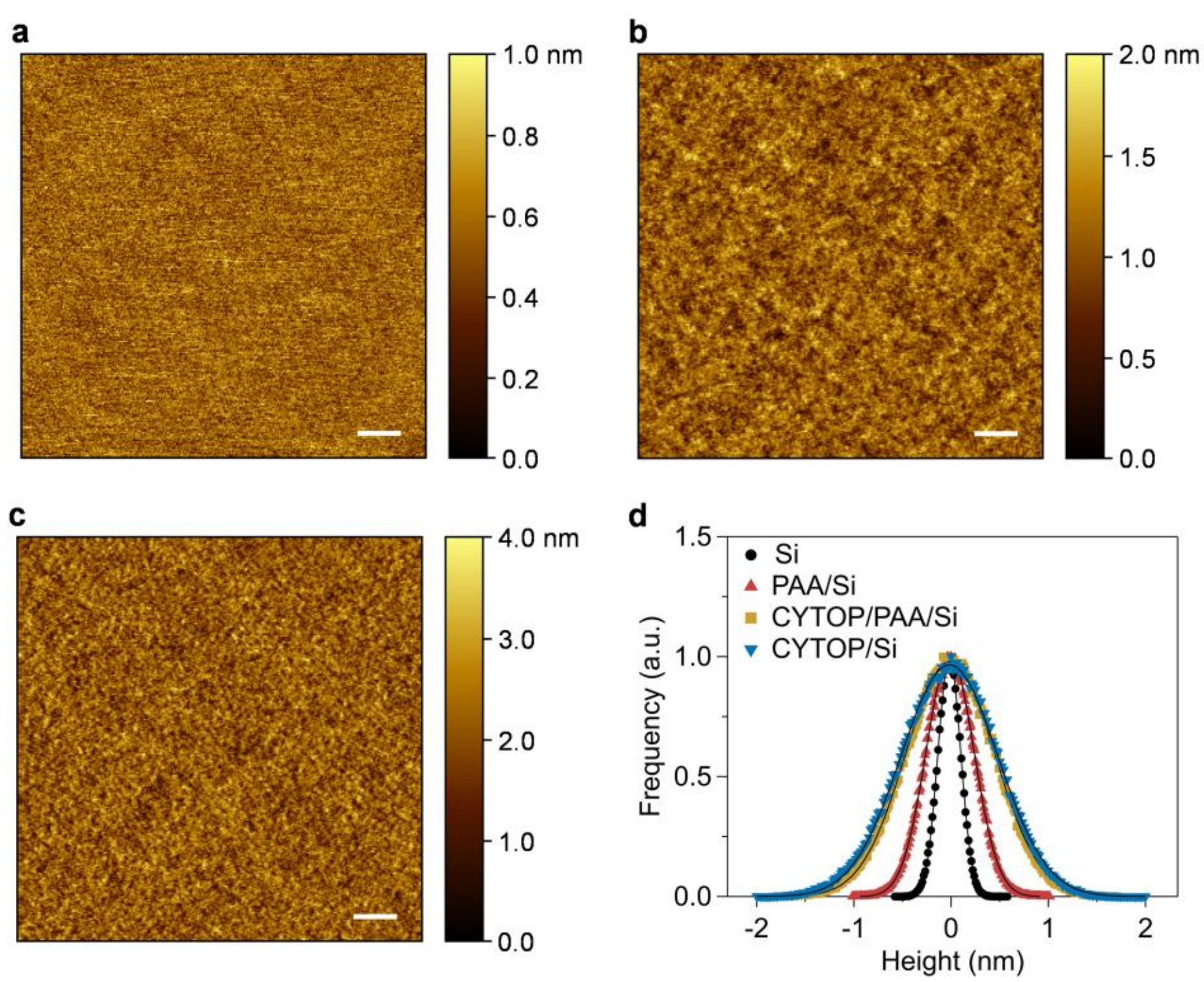

**Figure S2.** AFM images of (a) Si, (b) PAA/Si, (c) CYTOP/Si and (d) height histogram measured by AFM for Si, PAA/Si, CYTOP/PAA/Si and CYTOP/Si. Solid lines in (d) are Gaussian fits to the distributions. Scale bars, 2 μm.

Figure S2 presents atomic force microscopy (AFM) images of a bare Si substrate, PAA-coated Si, and CYTOP-coated Si without PAA. The AFM analysis of the bare Si substrate (Figure S2(a)) revealed a smooth and uniform surface morphology, with a root-mean-square (rms) roughness of 0.10 ± 0.01 nm, measured over a 20 μm × 20 μm scan area at 3 different locations. Figure S2(b) shows the AFM image of PAA coated on Si, which similarly exhibits a homogeneous and smooth surface, with rms roughness of 0.25 ± 0.02 nm. This indicates that the PAA layer maintains good uniformity without introducing significant topographical variations. AFM measurements of CYTOP spin-coated directly on Si (Figure S2(c)) reveal that the surface morphology remains nearly unchanged compared to CYTOP coated on PAA/Si (as shown in Figure 2(i) of the main manuscript). The rms roughness in this case is also comparable to that of CYTOP/PAA/Si, demonstrating that CYTOP forms a uniform and conformal film irrespective of the underlying substrate. Figure S2(d) presents histograms of the surface height distributions for the different samples, highlighting variations in roughness and providing a quantitative comparison of the topographical uniformity across Si, PAA/Si, and CYTOP films coated on Si and PAA/Si.

### 3. Fabrication Process of H-terminated Diamond FETs

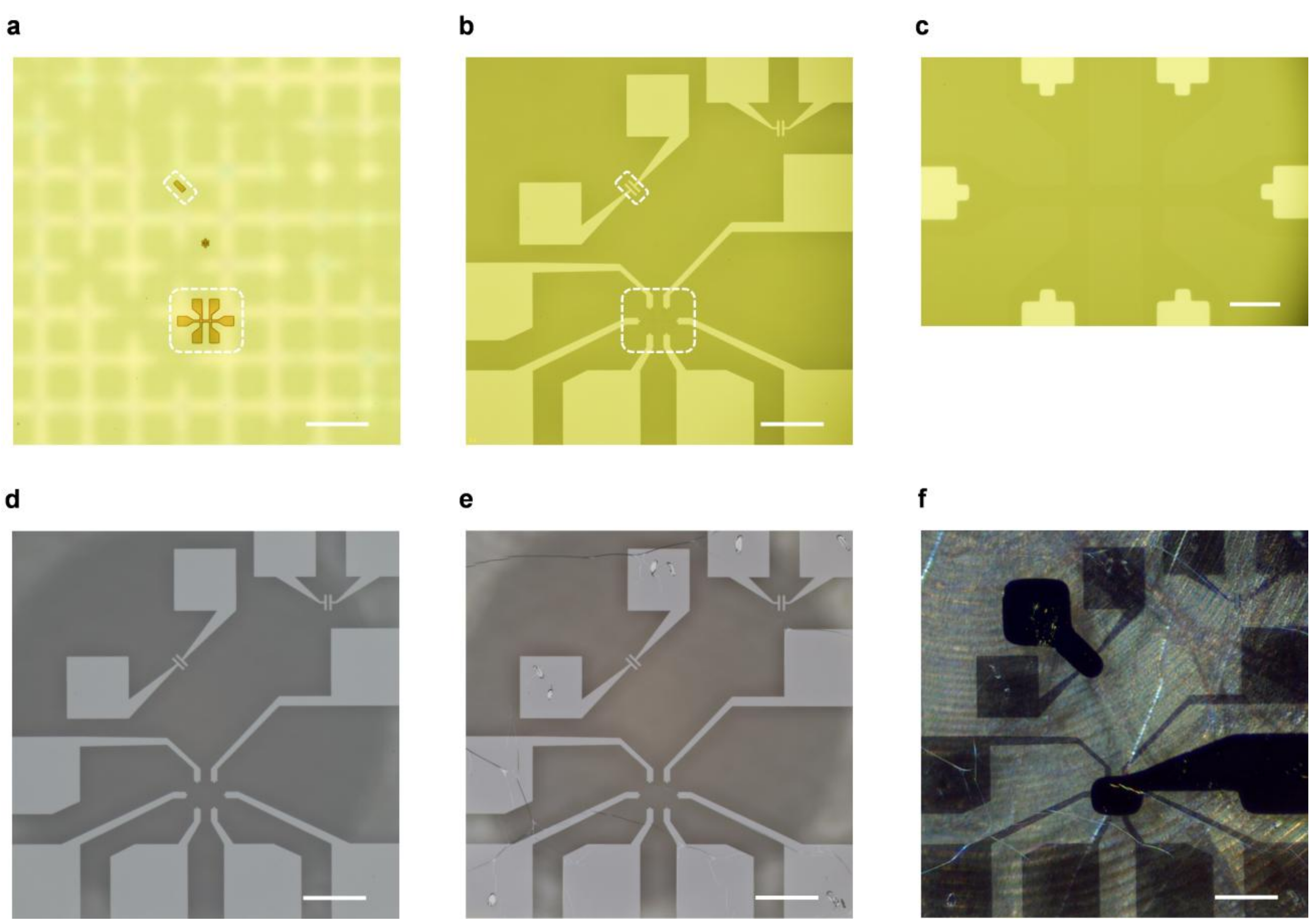

**Figure S3.** Optical microscope images showing the stepwise fabrication of the diamond FETs. (a) Undoped diamond substrate partially covered with photoresist mask for N-ion implantation. (b) Patterned electrodes on both unimplanted and implanted regions. (c) Magnified image displaying the unimplanted region with darker contrast. (d) Surface after H-termination. (e) After lamination of the CYTOP film. (f) Final FET structure after gate electrode deposition. All scale bars correspond to 200 μm, except for (c), which corresponds to 20 μm.

As discussed in the main manuscript, the lamination technique of the CYTOP fluoropolymer insulator was employed to fabricate FETs on H-terminated diamond substrates. Prior to the lamination process, lateral isolation of the active region was achieved via nitrogen-ion (N-ion) implantation, which acts as deep donor in diamond. This was performed by selectively covering the desired active region with photoresist mask, thereby shielding it from ion exposure during implantation. It has been shown[49,50] that the surface conductivity of H-terminated diamond can be effectively suppressed by implanting a sufficiently high concentration of nitrogen near the surface. This suppression arises because the surface Fermi level remains significantly above the valence band maximum,[49] owing to the deep donor levels of nitrogen, which hinder the generation of holes via the surface transfer doping mechanism.[46,48] Furthermore, the formation of conductive electrons in N-ion implanted diamond is also unlikely, since the donor levels associated with nitrogen are substantially deep. A comprehensive analysis of the electrical isolation achieved by N-ion implantation has been provided elsewhere.[50]

Figure S3(a) displays an optical microscope image of a diamond substrate patterned with photoresist mask to define the active region. The ion implantation was carried out using $^{15}N$ ions with a fluence of $1 \times 10^{14}$ $cm^{-2}$ at an energy of 10 keV under room temperature conditions. Following implantation, the photoresist mask was removed, and the substrates were subjected to a thorough cleaning process involving hot acid treatment and subsequent organic solvent cleaning to remove residual contaminants. Subsequently, bottom electrodes were defined on the cleaned substrates using maskless lithography followed by e-beam evaporation of Ti/Pt (5 nm/5 nm) metal electrodes, as shown in Figure S3(b). The electrodes were patterned in both Hall bar and two-probe geometries on the unimplanted (active) region of the substrate, as marked by dashed boxes in Figures S3(a) and S3(b). Figure S3(c) presents a magnified optical microscope image of the Hall bar geometry, displaying the contrast between the implanted and unimplanted regions. The darker region corresponds to the unimplanted, electrically active channel area where the electrodes are placed. Following N-ion implantation and electrode deposition, the diamond substrate was transferred to a microwave plasma-assisted chemical vapor deposition (MPCVD) system for global H-termination. The detailed procedure for H-termination and subsequent transfer of H-terminated diamond without air exposure is provided in the main manuscript. Figures S3(d) and S3(e) present optical microscope images of the H-terminated diamond and CYTOP-laminated H-terminated diamond, respectively, both captured inside an Ar-filled glove box. After lamination, an Al/Au (50 nm/100 nm) bilayer gate electrode was deposited using a shadow mask via thermal evaporation on both two-terminal and Hall bar configurations. The completed devices were then transferred to an $N_2$-filled glove box for electrical measurements. Figure S3(f) shows an optical microscope image of the fabricated FET captured inside the $N_2$-filled glove box using single layer CYTOP film.

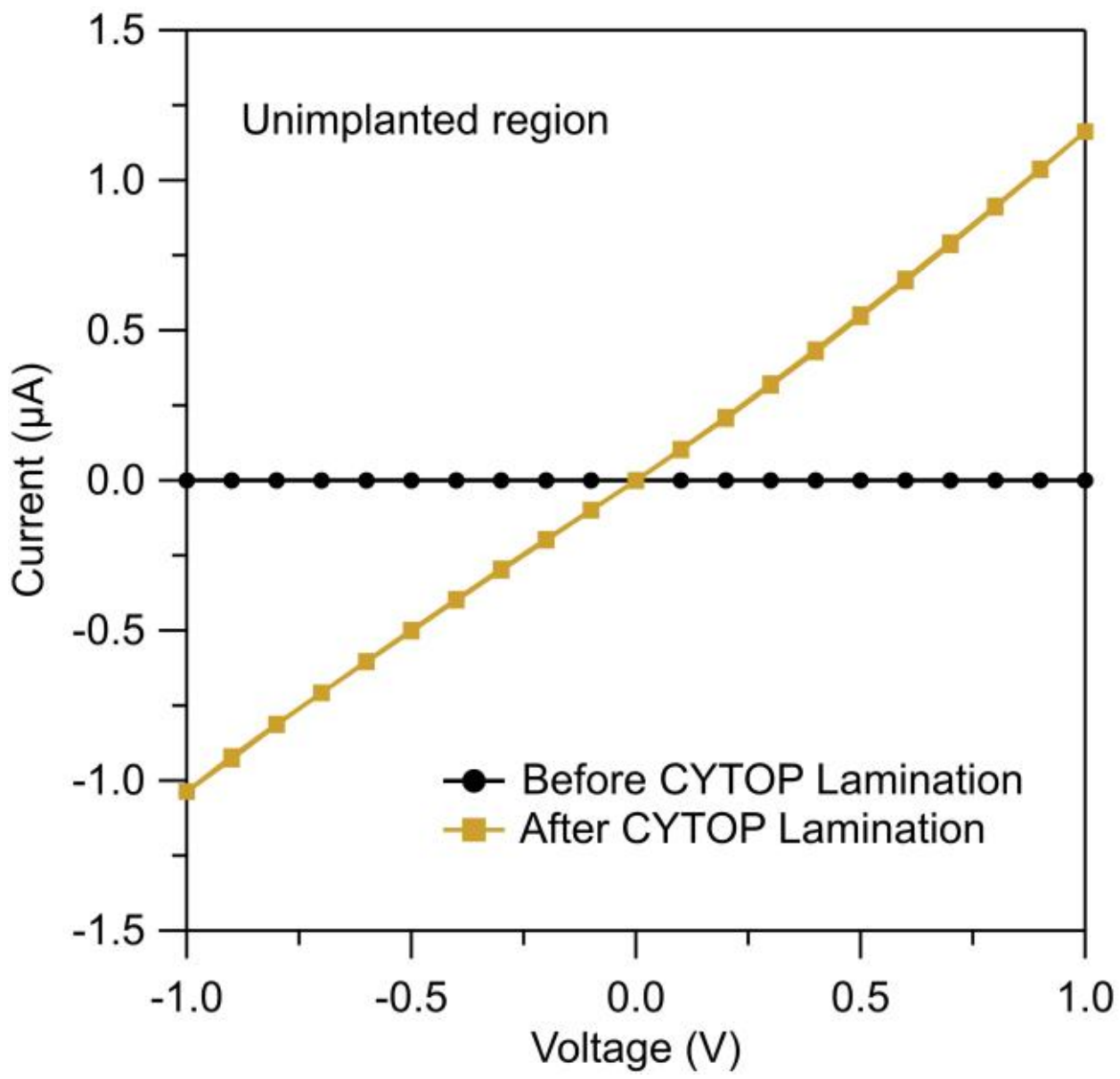

**Figure S4.** *I-V* characteristics of unimplanted region on H-terminated diamond before and after lamination with CYTOP gate insulator films without exposing H-terminated diamond to ambient air.

To assess the influence of CYTOP film on the electrical characteristics of H-terminated diamond, *I-V* measurements were conducted on both unimplanted and N-ion implanted regions. These measurements were performed in Ar-filled glove box to minimize the effect of ambient air on the H-terminated diamond surface. Figure S4 presents the *I-V* curves measured in two-probe geometry on unimplated region before and after the lamination of the CYTOP insulating film without exposing diamond surface to ambient air. These results correspond to the device fabricated by stacking two layers of CYTOP on H-terminated diamond. The *I-V* measurements after CYTOP film lamination were carried out before the deposition of gate electrode (*i.e.* at the stage shown in Figures S3(d) and S3(e)). The calculated resistance values were subsequently converted to sheet resistance by multiplying by the width of the unimplanted region (or electrode width for the implanted region) divided by the electrode spacing. The sheet resistance before lamination was estimated to be $6 \times 10^{11}$ Ω, confirming the high resistive nature of the H-terminated diamond surface, as diamond surface was not exposed to ambient air. After the CYTOP lamination, the sheet resistance dropped to $4.5 \times 10^6\ \Omega$, indicating the emergence of surface conductivity. This conductivity is attributed to the charge trapping sites within the CYTOP film, which can extract electrons from the valence band of H-terminated diamond and generate holes at the diamond surface, similar to the surface transfer doping mechanism reported for H-terminated diamond with deposited gate dielectrics.[48] Possible mechanisms responsible for the emergence of surface conductivity after CYTOP film lamination are discussed in the *Discussion Section* of the main manuscript. The *I-V* measurements on implanted regions were also carried out (not shown here), which showed nearly no change in the resistivity values before and after CYTOP lamination, revealing that good isolation was achieved by N-ion implantation in diamond, consistent with results reported elsewhere.[50]

**4. H-terminated Diamond FETs using Double-layer CYTOP Insulating Films**

As discussed in the main manuscript, the electrical performance of the fabricated diamond-based FETs was systematically evaluated at room temperature by measuring both output and transfer characteristics. To verify the reproducibility and reliability of our fabrication method, three FETs were prepared, two with single-layer CYTOP and one with double-layer CYTOP, with each layer thickness of 90 nm. Figure S5 shows the data of a *p*-channel H-terminated diamond FET fabricated by stacking two 90 nm thick CYTOP films (total thickness of 180 nm), while the data of an FET fabricated with a single-layer CYTOP film is presented in main manuscript. The geometrical parameters of the Hall-bar configuration, including the channel length ($L_G$), channel width ($W_G$), and the separation between voltage probes ($L_p$), were determined from optical microscope images. For this device, $L_G$ was 117.2 μm, $W_G$ was 7.9 μm, and $L_p$ was 29.6 μm. Gate-voltage-dependent Hall measurements were carried out to obtain mobility and sheet carrier density by sweeping the magnetic field between −0.5 and +0.5 T, and the results are presented respectively in Figures S5(a) and S5(b). It is seen that Hall mobility and carrier density values increase monotonously with increase of gate voltage, reaching the maximum values of 250 $cm^2V^{-1}s^{-1}$ and $1.1 \times 10^{12}\ cm^{-2}$, respectively, at gate voltage of −20 V. Figure S5(c) shows the sheet conductance measured in four-probe configuration by sweeping the gate voltage ($V_{GS}$) from +5 to −20 V and back from −20 to +5 V, while applying drain voltage ($V_{DS}$) of ±100 mV. The maximum sheet conductance at gate voltage of −20 V was observed to be $4.7 \times 10^{-5}\ \Omega^{-1}$, corresponding to a sheet resistance of 21 kΩ. By extrapolating the linear region of sheet conductance in the gate voltage range of −10 V to −20 V, the threshold voltage ($V_{th}$) was estimated to be −4.2 V.

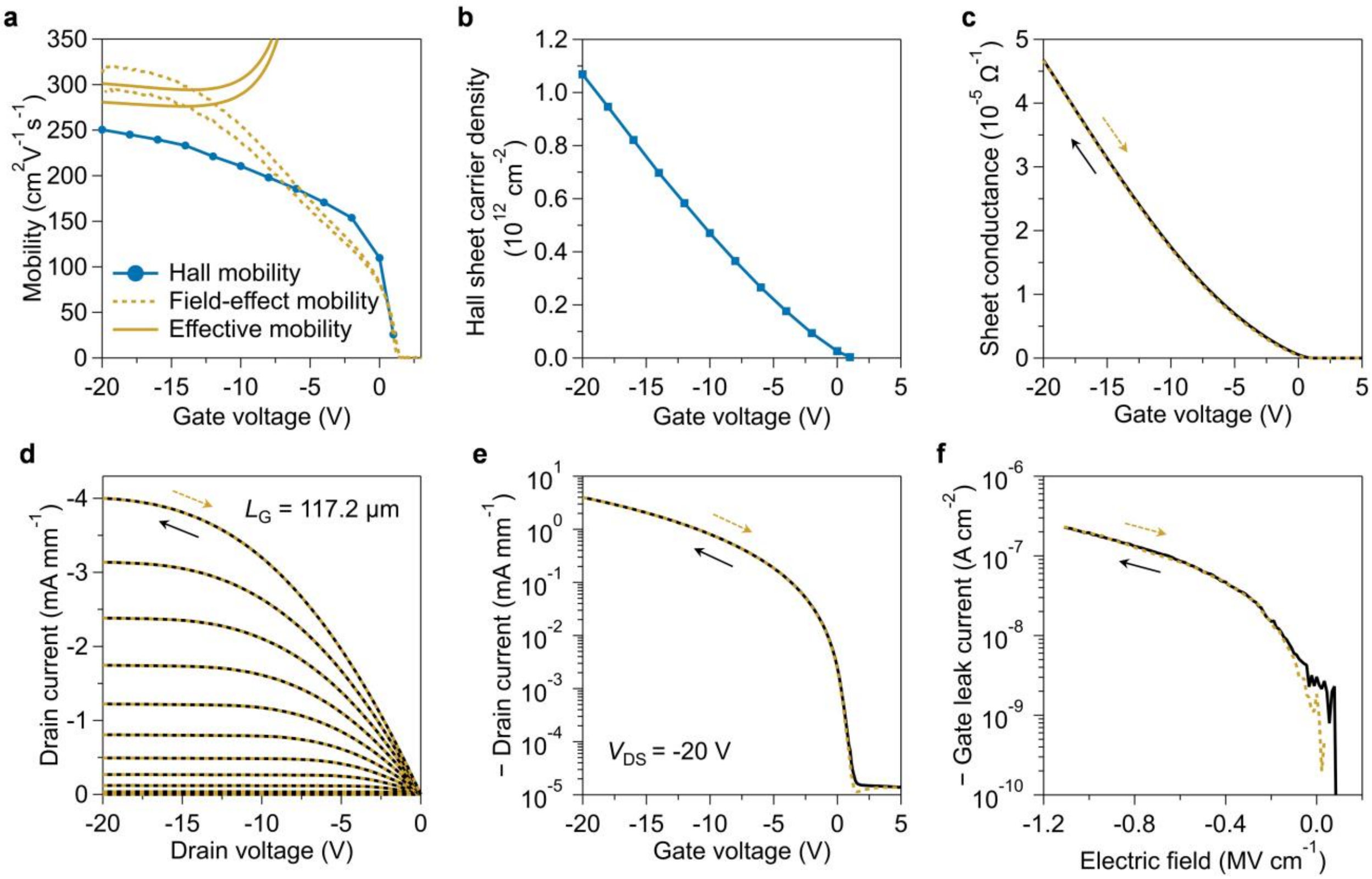

**Figure S5**. Electrical characteristics of the diamond FET with double-layer CYTOP insulator. (a) Gate voltage dependent Hall mobility, along with field-effect and effective mobilities. Among the two solid lines, the higher one corresponds to the effective mobility obtained without air exposure of FET, while the lower one corresponds to that after air exposure. Similarly, for the dashed lines, the higher one represents the field-effect mobility without air exposure, whereas the lower one corresponds to the values obtained after air exposure. (b) Hall sheet carrier density versus gate voltage. (c) Gate voltage dependence of sheet conductance in the linear region obtained from four-probe measurements, where the gate voltage was swept from +5 to −20 V and back. (d) Output characteristics measured using a two-probe configuration, showing the drain current versus drain voltage for gate voltages varied from +4 to −20 V in 2 V steps. (e) Drain current versus gate voltage at a fixed drain voltage of −20 V. (f) Gate leak current density versus electric field, where the electric field is calculated by dividing the gate voltage by the CYTOP film thickness. In (c)-(f), solid black and dashed brown lines correspond to forward and backward sweeps, respectively.

The effective mobility was calculated to be ≈300 $cm^2V^{-1}s^{-1}$ and maximum carrier density was estimated to be $9.8 \times 10^{11}$ $cm^{-2}$ at $V_{GS}$ of −20 V. Most of the FET measurements were carried out inside an $N_2$-filled glove box without exposing the device to air, whereas Hall measurements required device transfer, during which ambient air exposure was unavoidable. Accordingly, Figure S5(a) presents both field-effect and effective (calculated using a threshold voltage of −4.2 V) mobilities before and after air exposure, derived from Figure S5(c). The procedure for calculating effective and field-effect mobilities is described in the main manuscript. Upon exposure to air, the effective and field-effect mobilities show slight decrease (≈10% reduction), which is attributed to the marginal surface transfer doping caused by the adsorption of airborne acceptor-like species in the channel region. These adsorbates can increase Coulomb scattering at the surface, thereby reducing the carrier mobility.[39] Note that the apparent increase in effective mobility as $V_{GS}$ approaches $V_{th}$ arises because the denominator in the mobility extraction formula approaches zero (effective mobility $\propto 1/|V_{GS} - V_{th}|$),[39] leading to a diverging mobility value.

Figure S5(d) presents the output characteristics, in which the drain current ($I_{DS}$) was recorded by sweeping $V_{DS}$ from 0 to −20 V and then in the reverse direction, at different $V_{GS}$ ranging from +4 to −20 V in 2 V steps. The device clearly shows current modulation across both the linear and saturation regions, confirming effective field-effect control of two-dimensional hole gas (2DHG) at the diamond surface. The drain current was normalized by dividing the values with $W_G$. The maximum drain current in this case was observed to be −4 mA mm$^{-1}$ at $V_{DS} = V_{GS} = -20$ V. In the linear region (low $V_{DS}$), $I_{DS}$ increases linearly with $V_{DS}$, indicating the formation of good ohmic contacts at the source and drain electrodes. As $V_{DS}$ is increased further, the current reaches a saturation point and becomes nearly unchanged. Figure S5(e) shows the transfer curve, where drain current was measured in two-probe configuration as a function of gate voltage at a fixed $V_{DS}$ = −20 V. An on/off current ratio of $2.8 \times 10^5$ was extracted from Figure S5(e). It is seen that the output and transfer curves display negligible hysteresis, indicative of substantially low interface trap density. The normalized hysteresis, obtained from Figure S5(e) in the gate voltage range of 0.3 to −1 V, was found to be ≈30 mV (MV cm$^{-1}$)$^{-1}$, which is slightly larger than the value presented in the main manuscript; however, it remains of the same order as those reported for highly cleaned 2D semiconductor interfaces.[14,18] The subthreshold swing (SS) from Figure S5(e) was determined to be approximately 600 mV dec$^{-1}$. The interface trap density ($D_{it}$) can be estimated by following the equations[67]

$$\mathrm{SS} = \ln 10 \frac{\mathrm{d}V_{\mathrm{GS}}}{\mathrm{d}(\ln I_D)} = \ln 10 \frac{k_{\mathrm{B}}T}{e} \frac{C_{\mathrm{CYTOP}} + C_{\mathrm{depl}} + e^2 D_{\mathrm{it}}}{C_{\mathrm{CYTOP}}} \quad (1),$$

$$C_{\mathrm{depl}} = \sqrt{\frac{\epsilon_{\mathrm{diamond}}\, e\, N_{\mathrm{D}}}{2\phi_{\mathrm{S}}}} \quad (2),$$

where $k_B$ is the Boltzmann constant, $T$ is the absolute temperature, $e$ is the elementary charge, $C_{CYTOP}$ (= $\epsilon_{CYTOP}/t_{CYTOP}$) is the capacitance of CYTOP gate insulator, and $C_{depl}$ is the capacitance of the diamond depletion layer. Here, $\epsilon_{diamond}$ is the permittivity ($5.7\epsilon_0$) of diamond and $\phi_S$ is surface potential. Using a relative dielectric constant of 2.0 for CYTOP, subthreshold swing of 600 mV

dec$^{-1}$, and thickness of 180 nm, the interface trap density for this device was estimated to be ≤6 × $10^{11}$ cm$^{-2}$ eV$^{-1}$. In this calculation, we assumed $C_{depl}$ to be zero, yielding the maximum possible value of $D_{it}$. The diamond substrates used in this study were purchased from Element Six, which possess boron ($N_A$) and nitrogen ($N_D$) impurity concentrations of <0.5 ppm and <1 ppm, respectively. When the depletion capacitance is taken into account by assuming $N_D$ = 1 ppm and $\phi_S$ = 4 V,[39,49] Equation (2) gives $C_{depl} = 4 \times 10^{-2}$ µF cm$^{-2}$. Under these conditions, the extracted $D_{it}$ value decreases to ≈3 × $10^{11}$ cm$^{-2}$ eV$^{-1}$. Similarly, the interface trap density reported in the main manuscript was also calculated under the assumption that $C_{depl} = 0$. However, when $C_{depl} = 4 \times 10^{-2}$ µF cm$^{-2}$ is considered, the estimated trap density further decreases to approximately 1 × $10^{11}$ cm$^{-2}$ eV$^{-1}$ for the FET presented in the main manuscript. The trap density for the FET with double layer CYTOP is slightly higher than the FET fabricated with single-layer CYTOP, but it remains nearly comparable to or lower than most reported values for H-terminated diamond FETs. The increase in trap density in this case may be attributed to the slight inhomogeneity of H-termination of diamond surface as well as to defects or imperfections on the diamond or CYTOP surfaces introduced during the fabrication process. This increase in trap density likely contributed to the observed reduction in mobility compared to the device reported in the main manuscript.

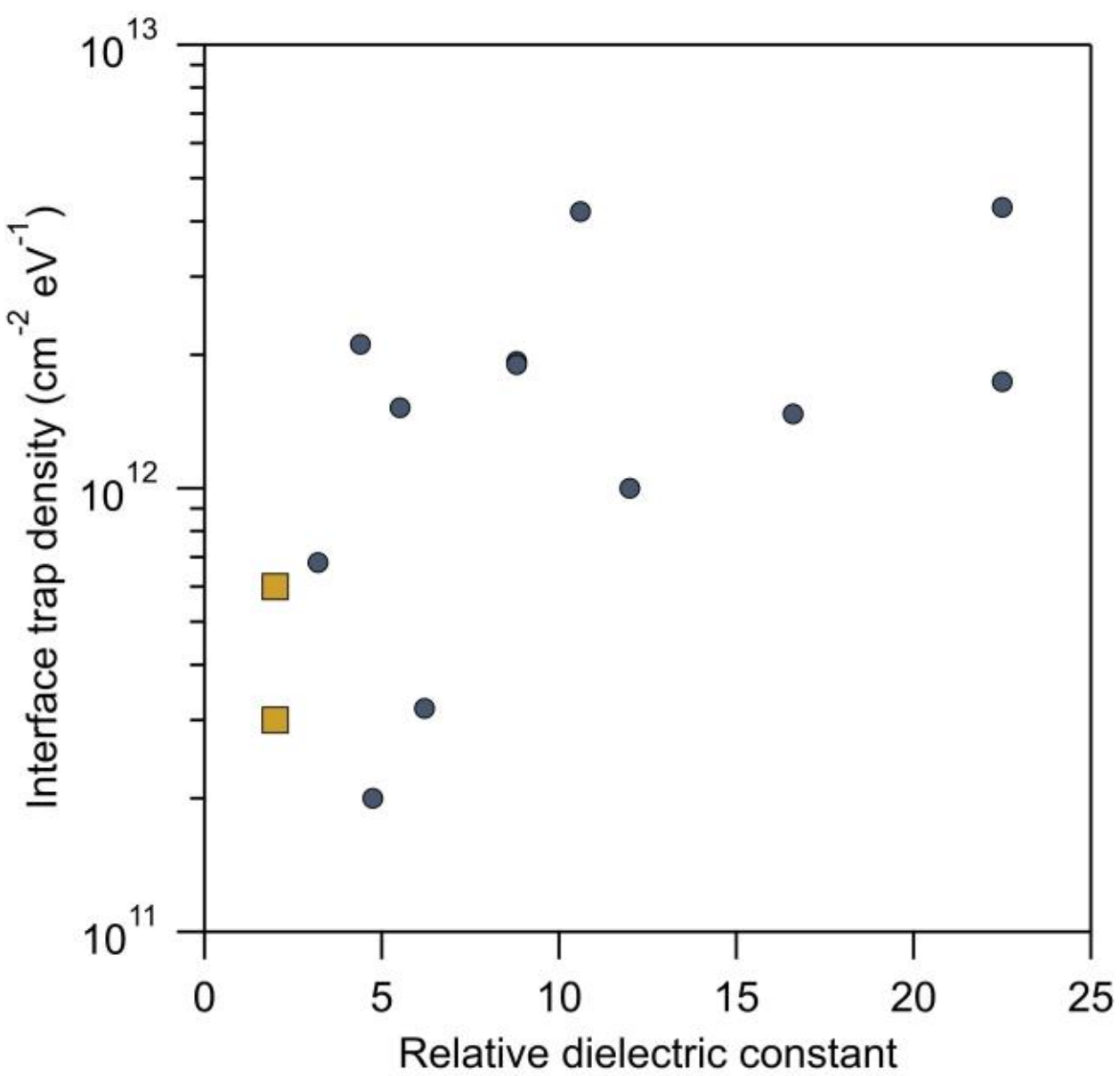

**Figure S6.** Comparison of the interface trap density versus relative dielectric constant for H-terminated diamond FETs. Yellow squares represent data from this work, while grey circles correspond to reported literature values.[39,45,52-54,68-73]

Figure S6 compares the interface trap density versus relative dielectric constant, including the highest $D_{it}$ values obtained in this work and extracted results from reported studies on H-terminated diamond FETs.[39,45,52-54,68-73] Figure S5(f) shows the gate leakage current density as a function of electric field. As also described in the main manuscript, the leakage current was measured by shorting the source and drain electrodes and sweeping $V_{GS}$ from +4 to −20 V and then in reverse direction. The leakage current was normalized to the active channel area of Hall bar under the gate electrode. Overall, the FET shows good device performance with substantially high hole mobility values highlighting the effectiveness and reproducibility of our device fabrication process.

## 5. Stability After Long-Term Storage

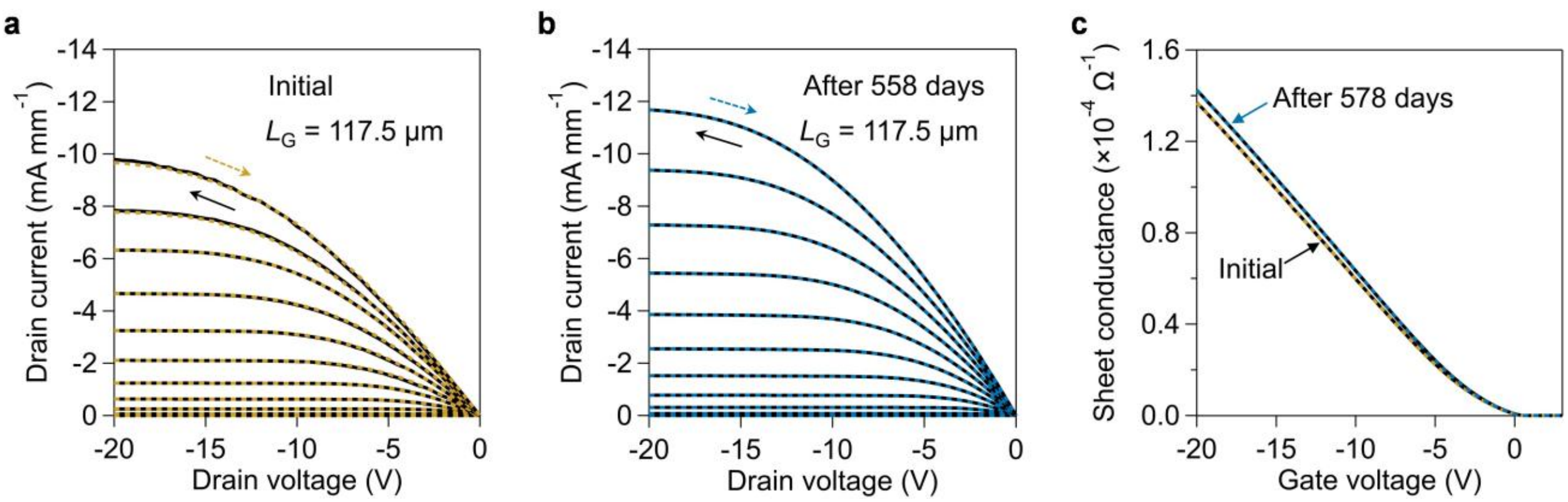

**Figure S7.** The stability of the diamond FET with CYTOP gate insulator after long-term storage. (a) Output characteristics measured immediately after device fabrication (as also shown in main manuscript) using a two-probe configuration, displaying the drain current versus drain voltage at different gate voltages of +2 to −20 V in 2 V steps. (b) Output characteristics of the same device measured after 558 days of storage under identical measurement conditions with gate voltages varied from +2 to −20 V in 2 V steps. (c) Comparison of the gate-voltage-dependent sheet conductance measured initially and after 578 days of storage in the linear region using a four-probe configuration while applying $V_{DS}$ of ±100 mV. Solid and dashed lines in (a)-(c) correspond to forward and backward sweeps, respectively.

The long-term stability of the device was assessed by comparing its electrical performance before and after extended storage. Following the initial measurements presented in the main manuscript, the same device was briefly exposed to ambient air for approximately one hour and then stored in an $N_2$-filled glove box. Figure S7(a) displays the output curves from the initial measurements, where the drain current is plotted against drain voltage ranging from 0 to -20 V at gate voltages from +2 to -20 V in 2 V steps. After 558 days of storage, output curves measured under similar conditions are shown in Figure S7(b). These results demonstrate a slight increase in drain current (by approximately 20%) compared to the initial values. Figure S7(c) presents the gate voltage dependence of the sheet conductance measured in a four-probe configuration, comparing the initial data with that obtained after 578 days of storage. The results exhibit no significant changes in either the conductance or threshold voltage, demonstrating excellent long-term electrical stability. Importantly, both measurements show negligible hysteresis, indicating stable dielectric and interface properties even after prolonged storage. The slight increase in drain current may be attributed to reduced contact resistance, likely arising from enhanced surface transfer doping near the electrodes as a result of ambient exposure. A minor interfacial gap between the electrodes and the CYTOP film may facilitate atmospheric adsorption, thereby reducing the contact resistance and increasing the drain current. Overall, these findings demonstrate that the transferred CYTOP dielectric film provides reliable gate control and excellent long-term device stability even after environmental exposure and extended storage.

## References

(1) Amano, H.; Baines, K.; Beam, E.; *et al*. The 2018 GaN power electronics roadmap. *J. Phys. D: Appl. Phys.* **2018**, *51*, 163001. https://doi.org/10.1088/1361-6463/aaaf9d

(2) Donato, N.; Rouger, N.; Pernot, J.; Longobardi, G.; Udrea, F. Diamond power devices: state of the art, modelling, figures of merit and future perspective. *J. Phys. D: Appl. Phys.* **2020**, *53*, 093001. https://doi.org/10.1088/1361-6463/ab4eab

(3) Kim, M. S.; Almuslem, A. S.; Babatain, W.; *et al*. Beyond Flexible: Unveiling the Next Era of Flexible Electronic Systems. *Adv. Mater.* **2024**, *36*, 2406424. https://doi.org/10.1002/adma.202406424

(4) Zeng, D.; Zhang, Z.; Xue, Z.; *et al*. Single-crystalline metal-oxide dielectrics for top-gate 2D transistors. *Nature* **2024**, *632*, 788. https://doi.org/10.1038/s41586-024-07786-2

(5) Murata, K.; Inoue, S.; Higashino, T.; Hasegawa, T. Impact of gate dielectric surfaces on carrier transport and injection in inverted-coplanar single-crystal organic transistors. *Phys. Rev. Appl*. **2025**, *24*, 024038. https://doi.org/10.1103/8stj-n7cf

(6) Kim, H. G.; Lee, H.-B. Atomic Layer Deposition on 2D Materials. *Chem. Mater.* **2017**, *29*, 3809. https://doi.org/10.1021/acs.chemmater.6b05103

(7) Wang, B.; Huang, W.; Chi, L.; Al Hashimi, M.; Marks, T. J.; Facchetti, A. High-k Gate Dielectrics for Emerging Flexible and Stretchable Electronics. *Chem. Rev.* **2018**, *118*, 5690-5754. https://doi.org/10.1021/acs.chemrev.8b00045

(8) Yang, S.; Liu, K.; Xu, Y.; Liu, L.; Li, H.; Zhai, T. Gate Dielectrics Integration for 2D Electronics: Challenges, Advances, and Outlook. *Adv. Mater.* **2023**, *35*, 2207901. https://doi.org/10.1002/adma.202207901

(9) Liu, H.; Xu, K.; Zhang, X.; Ye, P. D. The integration of high-$\kappa$ dielectric on two-dimensional crystals by atomic layer deposition. *Appl. Phys. Lett.* **2012**, *100*, 152115. https://doi.org/10.1063/1.3703595

(10) McDonnell, S.; Brennan, B.; Azcatl, A.; *et al*. $HfO_2$ on $MoS_2$ by Atomic Layer Deposition: Adsorption Mechanisms and Thickness Scalability. *ACS Nano* **2013**, *7*, 10354-10361. https://doi.org/10.1021/nn404775u

(11) Li, N.; Wei, Z.; Zhao, J.; *et al*. Atomic Layer Deposition of $Al_2O_3$ Directly on 2D Materials for High-Performance Electronics. *Adv. Mater. Interfaces* **2019**, *6*, 1802055. https://doi.org/10.1002/admi.201802055

(12) Nam, T.; Seo, S.; Kim, H. Atomic Layer Deposition of a Uniform Thin Film on Two-Dimensional Transition Metal Dichalcogenides. *J. Vac. Sci. Technol. A* **2020**, *38*, 030803. https://doi.org/10.1116/6.0000068

(13) Yin, Y.-T.; Huang, C.-C.; Chiu, P.-H.; Jiang, Y.-S.; Hoo, J.-Y.; Chen, M.-J. High-Quality $HfO_2$ High-$\kappa$ Gate Dielectrics Deposited on Highly Oriented Pyrolytic Graphite via Enhanced Precursor Atomic Layer Seeding. *ACS Appl. Electron. Mater.* **2025**, *7*, 1943-1952. https://doi.org/10.1021/acsaelm.4c02224

(14) Huang, J.-K.; Wan, Y.; Shi, J.; *et al*. High-$\kappa$ perovskite membranes as insulators for two-dimensional transistors. *Nature* **2022**, *605*, 262. https://doi.org/10.1038/s41586-022-04588-2

(15) Yang, A. J.; Han, K.; Huang, K.; *et al*. Van der Waals integration of high-$\kappa$ perovskite oxides and two-dimensional semiconductors. *Nat. Electron.* **2022**, *5*, 233-240. https://doi.org/10.1038/s41928-022-00753-7

(16) Lu, Z.; Chen, Y.; Dang, W.; *et al*. Wafer-scale high-$\kappa$ dielectrics for two-dimensional circuits via van der Waals integration. *Nat. Commun.* **2023**, *14*, 2340. https://doi.org/10.1038/s41467-023-37887-x

(17) Cong, X.; Gao, X.; Sun, H.; *et al*. Epitaxial integration of transferable high-$\kappa$ dielectric and 2D semiconductor. *J. Am. Chem. Soc.* **2024**, *146*, 20837. https://doi.org/10.1021/jacs.4c04984

(18) Yao, Z.; Tian, H.; Sasaki, U.; *et al*. Transferrable, wet-chemistry-derived high-$\kappa$ amorphous metal oxide dielectrics for two-dimensional electronic devices. *Nat. Commun.* **2025**, *16*, 1482. https://doi.org/10.1038/s41467-025-56815-9

(19) He, Y.; Lv, Z.; Liu, Z.; *et al*. Sacrifice-layer-free transfer of wafer-scale atomic-layer-deposited dielectrics and full-device stacks for two-dimensional electronics. *Nat. Commun.* **2025**, *16*, 5904. https://doi.org/10.1038/s41467-025-60864-5

(20) Lin, C.-Y.; Chen, B.-C.; Liu, Y.-C.; *et al*. Integration of freestanding hafnium zirconium oxide membranes into two-dimensional transistors as a high-$\kappa$ ferroelectric dielectric. *Nat. Electron.* **2025**, *8*, 560. https://doi.org/10.1038/s41928-025-01398-y

(21) AGC Chemicals, *CYTOP® Technical Brochure*, AGC Chemicals, **2023**. https://www.agc-chemicals.com/file.jsp?id=jp/en/fluorine/products/cytop/download/pdf/CYTOP_EN_Brochure.pdf

(22) Kalb, W. L.; Mathis, T.; Haas, S.; Stassen, A. F.; Batlogg, B. Organic small molecule field-effect transistors with Cytop™ gate dielectric: Eliminating gate bias stress effects. *Appl. Phys. Lett.* **2007**, *90*, 092104. https://doi.org/10.1063/1.2709894

(23) Wark, A. W.; Lee, H. J.; Corn, R. M. Long-Range Surface Plasmon Resonance Imaging for Bioaffinity Sensors. *Anal. Chem.* **2005**, *77*, 3904. https://doi.org/10.1021/ac050402v

(24) Hwang, D. K.; Fuentes-Hernandez, C.; Kim, J. B.; Potscavage, W. J., Jr.; Kippelen, B. Flexible and stable solution-processed organic field-effect transistors. *Org. Electron.* **2011**, *12*, 1108. https://doi.org/10.1016/j.orgel.2011.04.002

(25) Nugraha, M. I.; Häusermann, R.; Bisri, S. Z.; *et al*. High Mobility and Low Density of Trap States in Dual-Solid-Gated PbS Nanocrystal Field-Effect Transistors. *Adv. Mater.* **2015**, *27*, 2107. https://doi.org/10.1002/adma.201404495

(26) Zhang, H.; Masuda, A.; Kawakami, R.; *et al*. Fluoropolymer Nanosheet as a Wrapping Mount for High-Quality Tissue Imaging. *Adv. Mater.* **2017**, 29, 1703139. https://doi.org/10.1002/adma.201703139

(27) Li, H.; Geng, D.; Lei, Y.; *et al*. Recent Progress in Fluorinated Dielectric-Based Organic Field-Effect Transistors and Applications. *Adv. Sensor Res.* **2023**, *2*, 2300034. https://doi.org/10.1002/adsr.202300034

(28) Liu, B.; Zhao, W.; Ding, Z.; *et al*. Engineering Bandgaps of Monolayer $MoS_2$ and $WS_2$ on Fluoropolymer Substrates by Electrostatically Tuned Many-Body Effects. *Adv. Mater.* **2016**, *28*, 6457. https://doi.org/10.1002/adma.201504876

(29) Zhang, H.; Yan, Q.; Xu, Q.; Xiao, C.; Liang, X. A sacrificial layer strategy for photolithography on highly hydrophobic surface and its application for electrowetting devices. *Sci. Rep.* **2017**, *7*, 3983. https://doi.org/10.1038/s41598-017-04342-z

(30) Qiu, Y.; Yang, S.; Sheng, K.; Photolithographic Patterning of Cytop with Limited Contact Angle Degradation. *Micromachines* **2018**, *9*, 509. https://doi.org/10.3390/mi9100509

(31) Swift, T.; Swanson, L.; Geoghegan, M.; Rimmer, S. The pH-responsive behaviour of poly(acrylic acid) in aqueous solution is dependent on molar mass. *Soft Matter* **2016**, *12*, 2542. https://doi.org/10.1039/c5sm02693h

(32) Song, G.; Wang, Y.; Tan, D. Q. A review of surface roughness impact on dielectric film properties. *IET Nanodielectr.* **2022**, *5*, 1. https://doi.org/10.1049/nde2.12026

(33) Feng, Q.-K.; Pei, J.-Y.; Zhang, Y.-X.; *et al*. Achieving high insulating strength and energy storage properties of all-organic dielectric composites by surface morphology modification. *Compos. Sci. Technol.* **2022**, *226*, 109545. https://doi.org/10.1016/j.compscitech.2022.109545

(34) Maex, K.; Baklanov, M. R.; Shamiryan, D.; Iacopi, F.; Brongersma, S. H.; Yanovitskaya, Z. S. Low dielectric constant materials for microelectronics. *J. Appl. Phys.* **2003**, *93*, 8793-8841. https://doi.org/10.1063/1.1567460

(35) Fang, Q.; Yi, K.; Zhai, T.; *et al*. High-performance 2D electronic devices enabled by strong and tough two-dimensional polymer with ultra-low dielectric constant. *Nat. Commun.* **2024**, *15*, 10780. https://doi.org/10.1038/s41467-024-53935-6

(36) Cao, L.; Dong, M.; Guo, X.; Li, M. Recent Advances in Porous Low-k Materials for Integrated Circuits. *ACS Appl. Mater. Interfaces* **2025**, *17*, 43983-44010. https://doi.org/10.1021/acsami.5c07498

(37) Wort, J. H.; Balmer, R. S. Diamond as an Electronic Material. *Mater. Today* **2008**, *11*, 22. https://doi.org/10.1016/S1369-7021(07)70349-8

(38) Kawarada, H.; Aoki, M.; Ito, M. Enhancement mode metal-semiconductor field effect transistors using homoepitaxial diamonds. *Appl. Phys. Lett.* **1994**, *65*, 1563. https://doi.org/10.1063/1.112915

(39) Sasama, Y.; Kageura, T.; Imura, M.; Watanabe, K.; Taniguchi, T.; Uchihashi, T.; Takahide, Y. High-mobility *p*-channel wide-bandgap transistors based on hydrogen-terminated diamond/ hexagonal boron nitride heterostructures. *Nat. Electron.* **2022**, *5*, 37. https://doi.org/10.1038/s41928-021-00689-4

(40) Huang, Y.; Xiao, J.; Tao, R.; *et al*. High mobility hydrogen-terminated diamond FET with h-BN gate dielectric using pickup method. *Appl. Phys. Lett.* **2023**, *123*, 112103. https://doi.org/10.1063/5.0165596

(41) Kawarada, H. Diamond p-FETs using two-dimensional hole gas for high frequency and high voltage complementary circuits. *J. Phys. D: Appl. Phys.* **2023**, *56*, 053001. https://doi.org/10.1088/1361-6463/aca61c

(42) Qu, C.; Maini, I.; Guo, Q.; Stacey, A.; Moran, D. A. J. Extreme Enhancement-Mode Operation Accumulation Channel Hydrogen-Terminated Diamond FETs with $V_{th} < -6$ V and High on-Current. *Adv. Electron. Mater.* **2024**, *10*, 2400770. https://doi.org/10.1002/aelm.202400770

(43) Sasama, Y.; Iwasaki, T.; Monish, M.; Watanabe, K.; Taniguchi, T.; Takahide, Y. Self-aligned gate electrode for hydrogen-terminated diamond field-effect transistors with a hexagonal boron nitride gate insulator. *Appl. Phys. Lett.* **2024**, *125*, 092103. https://doi.org/10.1063/5.0224192

(44) Sasama, Y.; Iwasaki, T.; Imura, M.; Watanabe, K.; Taniguchi, T.; Takahide, Y. Enhanced channel mobility of hexagonal boron nitride/hydrogen-terminated diamond heterojunction field-effect transistor. *Appl. Phys. Lett.* **2025**, *127*, 143502. https://doi.org/10.1063/5.0272041

(45) Zhang, M.; Lin, F.; Wang, W.; *et al*. An enhancement-mode C-H diamond FET with low work function gate material gadolinia. *Appl. Phys. Lett.* **2025**, *126*, 132105. https://doi.org/10.1063/5.0250891

(46) Maier, F.; Riedel, M.; Mantel, B.; Ristein, J.; Ley, L. Origin of Surface Conductivity in Diamond. *Phys. Rev. Lett.* **2000**, *85*, 3472. https://doi.org/10.1103/PhysRevLett.85.3472

(47) Takagi, Y.; Shiraishi, K.; Kasu, M.; Sato, H. Mechanism of hole doping into hydrogen terminated diamond by the adsorption of inorganic molecule. *Surf. Sci.* **2013**, *609*, 203. https://doi.org/10.1016/j.susc.2012.12.015

(48) Crawford, K. G.; Maini, I.; Macdonald, D. A.; Moran, D. A. J. Surface transfer doping of diamond: A review. *Prog. Surf. Sci.* **2021**, *96*, 100613. https://doi.org/10.1016/j.progsurf.2021.100613

(49) Kageura, T.; Sasama, Y.; Yamada, K.; Kimura, K.; Onoda, S.; Takahide, Y. Surface transfer doping of hydrogen-terminated diamond probed by shallow nitrogen-vacancy centers. *Carbon* **2024**, *229*, 119404. https://doi.org/10.1016/j.carbon.2024.119404

(50) Hino, K.; Monish, M.; Sasama, Y.; Sakamoto, K.; Takahide, Y. Effective lateral isolation for hydrogen-terminated diamond field-effect transistors via nitrogen ion implantation. *Appl. Phys. Lett.* **2026**, *128*, 213503. https://doi.org/10.1063/5.0323938

(51) Zhang, M.; Wang, W.; Chen, G.; *et al*. Normally OFF Hydrogen-Terminated Diamond Field-Effect Transistor With $Ti/TiO_x$ Gate Materials. *IEEE Trans. Electron Devices* **2020**, *67*, 4784. https://doi.org/10.1109/TED.2020.3025515

(52) Su, J.; Wang, W.; Shao, G.; Chen, G.; Wang, H.-X. Mobility-Enhanced Normally Off Hydrogen-Terminated Diamond FET With Low Interface State Density Using $Al_2O_3$/Nd Gate Stack. *Appl. Phys. Lett.* **2023**, 123, 172105. https://doi.org/10.1063/5.0171832

(53) Liang, Y.; Wang, W.; Niu, T.; Chen, G.; Wang, F.; He, S.; Zhang, M.; Wang, Y.; Wen, F.; Wang, H.-X. Very low subthreshold swing normally-off diamond FET and its logic inverters. *Appl. Phys. Lett.* **2025**, *125*, 263501. https://doi.org/10.1063/5.0221365

(54) Liang, Y.; Wang, W.; Niu, T.; *et al*. Normally-Off High-Performance Diamond FET with Large $V_{TH}$ and Low Leakage Current. *IEEE Trans. Electron Devices* **2025**, *72*, 12-16. https://doi.org/10.1109/TED.2024.3496447

(55) Liu, J. W.; Liao, M. Y.; Imura, M.; Koide, Y. Normally-off $HfO_2$-gated diamond field effect transistors. *Appl. Phys. Lett.* **2013**, *103*, 092905. https://doi.org/10.1063/1.4820143

(56) Liu, J. W.; Oosato, H.; Liao, M. Y.; Koide, Y. Enhancement-mode hydrogenated diamond metal-oxide semiconductor field-effect transistors with $Y_2O_3$ oxide insulator grown by electron beam evaporator. *Appl. Phys. Lett.* **2017**, *110*, 203502. https://doi.org/10.1063/1.4983091

(57) Oi, N.; Kudo, T.; Inaba, M.; *et al*. Normally-OFF Two-Dimensional Hole Gas Diamond MOSFETs Through Nitrogen-Ion Implantation. *IEEE Electron Device Lett.* **2019**, *40*, 933. https://doi.org/10.1109/LED.2019.2912211

(58) Ren, Z.; Xing, Y.; Lv, D.; *et al*. H-diamond MOS interface properties and FET characteristics with high-temperature ALD-grown $HfO_2$ dielectric. *AIP Adv.* **2021**, 11, 035041. https://doi.org/10.1063/5.0044004

(59) Ma, Z.-C.; Gao, N.; Cheng, S.-H.; *et al*. Surface Oxygen Adsorption and Electric Property of Hydrogen-Terminated Single Crystal Diamonds by UV/ozone Treatment. *Chin. Phys. Lett.* **2020**, *37*, 046801. https://doi.org/10.1088/0256-307X/37/6/066801

(60) Kageura, T.; Sasama, Y.; Teraji, T.; *et al*. Spin-State Control of Shallow Single NV Centers in Hydrogen-Terminated Diamond. *ACS Appl. Mater. Interfaces* **2024**, *16*, 13212. https://doi.org/10.1021/acsami.3c17544

(61) Kim, S.; Suzuki, K.; Sugie, A.; Yoshida, H.; Yoshida, M.; Suzuki, Y. Effect of end group of amorphous perfluoro-polymer electrets on electron trapping. *Sci. Technol. Adv. Mater.* **2018**, *19*, 486. https://doi.org/10.1080/14686996.2018.1477395

(62) Maier, F.; Ristein, J.; Ley, L. Electron affinity of plasma-hydrogenated and chemically oxidized diamond (100) surfaces. *Phys. Rev. B* **2001**, *64*, 165411. https://doi.org/10.1103/PhysRevB.64.165411

(63) Daicho, A.; Saito, T.; Kurihara, S.; Hiraiwa, A.; Kawarada, H. High-reliability passivation of hydrogen-terminated diamond surface by atomic layer deposition of $Al_2O_3$. *J. Appl. Phys.* **2014**, *115*, 223711. https://doi.org/10.1063/1.4881524

(64) Takada, T.; Kikuchi, H.; Miyake, H.; Tanaka, Y.; Yoshida, M.; Hayase, Y. Determination of charge-trapping sites in saturated and aromatic polymers by quantum chemical calculation. *IEEE Trans. Dielectr. Electr. Insul.* **2015**, *22*, 1240. https://doi.org/10.1109/TDEI.2015.7076827

(65) Kasu, M.; Saha, N. C.; Oishi, T.; Kim, S.-W. Fabrication of diamond modulation-doped FETs by $NO_2$ delta doping in an $Al_2O_3$ gate layer. *Appl. Phys. Express* **2021**, *14*, 051004. https://doi.org/10.35848/1882-0786/abf445

(66) Yang, Y.; Koeck, F. A.; Wang, X.; Nemanich, R. J. Surface transfer doping of $MoO_3$ on hydrogen terminated diamond with an $Al_2O_3$ interfacial layer. *Appl. Phys. Lett.* **2022**, *120*, 191602. https://doi.org/10.1063/5.0083971

(67) Sze, S. M.; Ng, K. K. *Physics of Semiconductor Devices*, 3rd ed. (John Wiley & Sons, 2007). ISBN:9780470068328

(68) Zhao, J.; Liu, J.; Sang, L.; Liao, M.; Coathup, D.; Imura, M.; Shi, B.; Gu, C.; Koide, Y.; Ye, H. Assembly of a high-dielectric constant thin $TiO_x$ layer directly on H-terminated semiconductor diamond. *Appl. Phys. Lett.* **2016**, *108*, 012105. https://doi.org/10.1063/1.4939650

(69) Liu, J. W.; Liao, M. Y.; Imura, M.; Banal, R. G.; Koide, Y. Deposition of $TiO_2/Al_2O_3$ bilayer on hydrogenated diamond for electronic devices: Capacitors, field-effect transistors, and logic inverters. *J. Appl. Phys.* **2017**, *121*, 224502. https://doi.org/10.1063/1.4985066

(70) Saha, N. C.; Kasu, M. Heterointerface properties of diamond MOS structures studied using capacitance-voltage and conductance-frequency measurements. *Diam. Relat. Mater.* **2019**, *91*, 219. https://doi.org/10.1016/j.diamond.2018.11.019

(71) Wang, W.; Wang, Y.; Zhang, M., *et al*. An Enhancement-Mode Hydrogen-Terminated Diamond Field-Effect Transistor With Lanthanum Hexaboride Gate Material. *IEEE Electron Device Lett.* **2020**, *41*, 585. https://doi.org/10.1109/LED.2020.2972330

(72) He, Q.; Zhang, J.; Ren, Z.; Zhang, J.; Su, K.; Lei, Y.; Lv, D.; Mi, T.; Hao, Y. Normally-off polycrystalline C-H diamond MISFETs with $MgF_2$ gate insulator and passivation. *Diam. Relat. Mater.* **2021**, *119*, 108547. https://doi.org/10.1016/j.diamond.2021.108547

(73) Zhang, M.; Wang, W.; Chen, G., *et al*. Electrical properties of cerium hexaboride gate hydrogen-terminated diamond field effect transistor with normally-off characteristics. *Carbon* **2023**, *201*, 71. https://doi.org/10.1016/j.carbon.2022.08.056